\documentclass[aps,prl,reprint,superscriptaddress,citeautoscript,floatfix,longbibliography]{revtex4-1}
\usepackage{graphicx}
\usepackage{epstopdf}
\usepackage{hyperref}
\usepackage{float}
\usepackage{amsmath}
\usepackage[export]{adjustbox}
\usepackage{xcolor}
\usepackage{cancel}
\usepackage{enumitem}
\usepackage{gensymb}
\usepackage{multirow}
\usepackage{array}

\usepackage[labelsep=period]{caption}
\usepackage[compact]{titlesec}
 \titlespacing{\section}{0pt}{2ex}{1ex}
 \titlespacing{\subsection}{0pt}{1ex}{1ex}
 \titlespacing{\subsubsection}{0pt}{0.5ex}{1ex}
 
\usepackage{cellspace}
\begin{document}

\title{Time-dependent Ginzburg-Landau treatment of RF Magnetic Vortices in Superconductors: Vortex-Semiloops in a Spatially Nonuniform Magnetic Field }

\author{Bakhrom Oripov}
\email[Corresponding author:]{bakhromtjk@gmail.com}
\affiliation{Quantum Materials Center, Department of Physics, University of Maryland, College Park, MD 20742, USA}

\author{Steven M. Anlage}
\affiliation{Quantum Materials Center, Department of Physics, University of Maryland, College Park, MD 20742, USA}

\date{\today}

\begin{abstract}
We apply time-dependent Ginzburg Landau (TDGL) numerical simulations to study the finite frequency electrodynamics of superconductors subjected to intense rf magnetic field. Much recent TDGL work has focused on spatially uniform external magnetic field and largely ignores the Meissner state screening response of the superconductor. In this work, we solve the TGDL equations for a \textit{spatially non-uniform magnetic field created} by a point magnetic dipole in the vicinity of a semi-infinite superconductor. A novel two-domain simulation is performed to accurately capture the effect of the inhomogeneous applied fields and the resulting screening currents. The creation and dynamics of vortex semiloops penetrating deep into the superconductor domain is observed and studied, and the resulting third-harmonic nonlinear response of the sample is calculated. The effect of point-like defects on vortex semi-loop behaviour is also studied. This simulation method will assist our understanding of the limits of superconducting response to intense rf magnetic fields.
\end{abstract}

\maketitle

\section{Introduction} \label{Intro}

Superconductor technology is widely used in industrial applications where high current and low loss are required. With technological advancements in the fabrication of high quality superconducting materials and significant reduction in cryocooler prices, superconductor-enabled devices like Magnetic Resonance Imaging (MRI), high performance microwave and radio frequency (RF) filters, low-noise and quantum-limited amplifiers, or fast digital circuits based on rapid single flux quantum (RSFQ) logic devices became feasible \cite{Seidel2015Book, SC_app}. \par

Superconducting Radio Frequency (SRF) cavities \cite{Aune2000TeslaCavity, Padamsee2008book} used in new generation high energy particle accelerators is an example of the large scale usage of superconductor technology. Nb is the most dominant material used in SRF applications because it has the highest superconducting critical temperature ($T_c=9.3K$) and super-heating field ($B_{sh}\approx240mT$) among the elemental superconductors at ambient pressure while being a good heat conductor at typical SRF operating temperatures \cite{Singer2015}. During normal operation an SRF cavity is subjected to high rf magnetic field parallel to the internal superconducting surface. One of the key objectives in SRF cavity operation is to maximize the accelerating gradient of the machine while minimizing the dissipated power in the cavities. However, these cavities remain susceptible to a number of issues including enhanced losses due to trapped magnetic flux \cite{Martinello2015} and the existence of the point-like surface defects \cite{Cooley2011, Iwashita2008}. The maximum gradient operating conditions are often limited by extrinsic problems. One limiting scenario is that a surface defect can facilitate the entrance of vortex-semiloops which can later be trapped due to the impurities within the bulk of the cavity \cite{Gurevich2008}. The energy dissipated due to the dynamics of these vortex semiloops under the influence of rf currents could be the limiting factor on the ultimate performance of the SRF cavity. This phenomenon cannot be simulated unless the effects of the screening currents and fields on the superconducting order parameter are included self-consistently. \par

This work is motivated by results for third-harmonic generation from a near-field microwave microscope utilized on Nb surfaces \cite{Lee2005GB, Mircea2009, Tai2011, Tai2013, Tai2014, Tai2015, Oripov2019}. In this experiment, a magnetic writer probe from a conventional magnetic recording hard-disk drive is used to create a high-intensity, localized, and inhomogeneous rf magnetic field on the surface of a Nb superconducting sample. This probe applies a localized field oscillating at microwave frequency, and measures the sample's fundamental \cite{Tai2014a} and harmonic rf response. In the experiment, the third-harmonic response and its dependence on the applied rf magnetic field amplitude and the temperature of the sample was studied. Preliminary results of TDGL modeling and comparison to experimental data was published in \cite{Oripov2019}. \par

In this work, numerical solutions of the time-dependent Ginzburg Landau (TDGL) equations are obtained for a superconductor subjected to a spatially nonuniform applied rf magnetic field, and the effect of boundary conditions on the accuracy of the results is investigated. First, the Ginzburg-Landau (GL) theory and its range of validity is discussed in section \ref{GLsec}. Secondly, the TDGL equations and the normalization used in this paper are presented in detail in section \ref{TDGLsec}. Then, the implementation of the TDGL simulation in COMSOL Multiphysics simulation software with all appropriate boundary conditions is summarized in section \ref{ComsolSec}. Later, in section IV.1 a two-domain simulation capable of correctly modeling spatially nonuniform magnetic fields and the response screening currents of the superconductor is described, and simple examples are presented to demonstrate the validity of the two-domain model. Next, in section V, an application of the two-domain simulation is presented, where vortex-semiloops created by a strongly inhomogeneous field distribution are simulated. The time evolution of the vortex-semiloops, their dependence on the magnitude of the rf magnetic field, and their interaction with a localized defect is studied. Finally, in section V.4, a more general case where the vortex-semiloops are created in a superconductor surface when a uniform rf magnetic field is applied parallel to the surface of the superconductor is presented, and the results are discussed. We then discuss future work in section VI and conclude the paper. \par

\section{Ginzburg-Landau Theory} \label{GLsec}

The Ginzburg-Landau (GL) theory is a generic macroscopic model appropriate for understanding the electrodynamic response of superconductors subjected to static magnetic fields and currents in the limit of weak superconductivity \cite{Landau1950}. GL generalizes the theory of superconductivity beyond BCS by explicitly considering inhomogeneous materials, including surfaces, interfaces, defects, vortices, etc. \par
The GL equations are differential equations which relate the spatial variation of the order parameter $\Psi(\vec{r})$ to the magnetic vector potential $\vec{A}(\vec{r})$ and the current $\vec{J}(\vec{r})$ in a superconductor. GL starts with an expression for the free energy density of a superconductor in terms of position-dependent order parameter and vector potential \cite{Ginzburg2009}:

\begin{widetext}
\begin{equation}
F_{GL}(\vec{r})=\alpha(\vec{r},T)\Psi^2 + \frac{\beta(\vec{r},T)}{2}\Psi^4+\cfrac{\hbar^2}{2m_*}\left \vert \left( \vec{\nabla}-\cfrac{ie_*}{\hbar}\vec{A} \right)\Psi \right \vert^2+\cfrac{1}{2\mu_0} \left \vert \vec{\nabla} \times \vec{A} - \vec{B}_a \right \vert^2
\label{F_GL}
\end{equation}
\end{widetext}

Here, $\alpha(\vec{r},T)$ and $\beta(\vec{r},T)$ are the temperature and position-dependent phenomenological expansion parameters, $m_*=2m_e$ is the mass of the Cooper pair, $e_*=2e$ is the charge of the Cooper pair, $\vec{B}_a=\vec{B}_a(\vec{r},t)$ is the amplitude of the externally applied magnetic field, and $i=\sqrt{-1}$. \par
Taking variational derivatives and minimizing the free energy with respect to $\Psi$ and $\vec{A}$ leads to the coupled Ginzburg-Landau equations \cite{Cyrot1973, Tinkham2004Book}:

\begin{equation}
\alpha(\vec{r},T) \Psi+\beta(\vec{r},T) \left \vert\Psi\right \vert^2\Psi+\cfrac{1}{2m_*} \left(\cfrac{\hbar}{i} \vec{\nabla} - e_*\vec{A} \right)^2 \Psi=0
\label{GL1}
\end{equation}

\begin{equation}
\cfrac{1}{\mu_0} \vec{\nabla} \times \left ( \vec{\nabla}\times \vec{A} - \vec{B}_a \right ) = \cfrac{e_* \hbar}{2m_*i} \left( \Psi^* \vec{\nabla} \Psi - \Psi \vec{\nabla} \Psi^* \right) - \cfrac{e_*^2}{m_*} \left \vert \Psi \right \vert^2 \vec{A}
\label{GL2}
\end{equation}

Apart from the TDGL model, the Bogoliubov-de Gennes (BdG) equations \cite{Bogoliubov1958_1, Bogoliubov1958_3, Gennes1964, Gennes1999Book}, Gorkov's Green function method \cite{Abrikosov1975book, Kopnin2001book, Gorkov1958}, the Matsubara formalism \cite{Eilenberger1968, SolidStatePhysics1983book} or Usadel's equations \cite{Usadel1970} can be used to study inhomogeneous superconductors \cite{Flokstra2010Thesis}. We shall utilize TDGL because of its relative simplicity and the physical insights it offers compared to these other more microscopic approaches.

\section{Time-Dependent Ginzburg Landau (TDGL) Equations and Normalization} \label{TDGLsec}

The GL equations are static, thus cannot be used to study the temporal evolution of the order parameter and the screening currents. In 1966, Schmid proposed a time-dependent generalization of the GL equations that could be utilized to study the dynamics of the order parameter \cite{Schmid1966}. Gor'kov and Eliashberg derived a similar equation \cite{Gorkov1968}, but noted that for the case of a gapped superconductor, there exists a singularity in the density of states vs energy spectrum which prohibits expanding various quantities in powers of the gap $\Delta$. \par

Gor'kov limited the use of TDGL to gapless superconductors, or to materials with magnetic impurities or other pair-breaking mechanisms that would round off the singularity in the BCS density of states \cite{Tinkham2004Book}. Proximity to a boundary with a normal metal, along with strong external magnetic fields and currents can also lead to gapless superconductivity before completely destroying it. Of relevance to the case of SRF cavities, numerous researchers have noted a substantial reduction in the singularity, and broadening of the density of states spectrum, under SRF operating conditions \cite{Gurevich2017SurfOptimize, Kubo2019} or with various types of impurities and imperfections at the surface \cite{Dynes1978, Dynes1984, Zasadzinski2003book, Balatsky2006, Proslier2008}. Such conditions would also justify the use and relevance of the TDGL equations under these circumstances.\par

In order to extend the validity of the TDGL formalism to gapped superconductors, a generalized version of TDGL (gTDGL) was proposed \cite{Kramer1978, Vodolazov2005}. In gTDGL, the effects of a finite inelastic electron scattering time are considered. gTDGL is valid for a superconductor in the dirty limit, but does not require strong limitations such as a large concentration of magnetic impurities and/or gapless superconductivity \cite{Kopnin2001book}. Nevertheless, both TDGL and gTDGL are not microscopic theories, thus some of the parameters of the model are difficult to determine precisely for a given material of interest . For this reason we focus on semi-quantitative results and use the phenomenological TDGL equations mainly to give insight into the signals created by our near-field microwave microscope \cite{Oripov2019}. Future work will explore the order parameter dynamics under gTDGL. In addition, questions of validity and relevance of the solution to the TDGL equations outside of the range in which they are derived remain. \par

 \par

 \par

TDGL numerical simulations have been employed on a broad variety of problems \cite{Cyrot1973, Vinokur1994, Aranson2002Review}. We note that TDGL was previously used to study vortex dynamics and V-I characteristics of 2-D rectangular thin films \cite{Machida1993}, vortex entry in the presence of twin boundaries \cite{Vinokur1996} and the vortex dynamics under ac magnetic field in mesoscopic superconductors \cite{Hernandez2008}. TDGL was also used to study the dynamics of vortex loops created by a \textit{static} magnetic dipole \cite{Berdiyorov2013} which is similar to the results discussed in this work. More recently, the TDGL formalism was used to estimate the strength of the Kerr effect in a superconductor when a short light pulse is applied \cite{Robson2017Kerr}. \par

Often a three dimensional problem is simplified by assuming that the sample is infinite in the direction parallel to the externally applied magnetic field, thus reducing the 3D problem to a 2D one \cite{Vinokur1996, Sardella2006, Sorensen2011}. Moreover, much published work done using numerical solutions to the TDGL equations involve problems with a spatially uniform external magnetic field and use a single (entirely superconducting) domain for the simulation. However, this assumption ignores the effect that the screening currents would have at the surface, which is one of the most important aspects of the problem that we investigate. \par

Here we give a brief motivation for the origins of the TDGL equations. Once the GL free energy is known in its functional form (Eq. \ref{F_GL}), the relaxation dynamical equation can be written by considering how the order parameter evolves after being slightly disturbed from its equilibrium value \cite{Kopnin2001book, Rogalla2011book}: 

\begin{equation}
-\gamma \left(\cfrac{\partial}{\partial t}+\cfrac{i}{\hbar}e_*\Phi \right)\Psi\left ( \vec{r},t \right ) = \cfrac{\delta F_{GL}(\vec{r,t})}{\delta \Psi^{*}}
\label{gamma}
\end{equation}

where $\gamma$ plays the role of a friction coefficient. Here, the scalar electric potential $\Phi$ is included to make the equation describing the dynamics of the superconducting order parameter gauge invariant. The TDGL equations are then derived through the variational derivatives of the GL free energy equation (Eq. \ref{F_GL}) with respect to $\Psi^*$ and $A$ and are given as follows \cite{Schmid1966, Hernandez2002, Sorensen2011}:

\begin{widetext}
\begin{equation}
\cfrac{\hbar^2}{2m_*D} \left(\cfrac{\partial}{\partial t}+\cfrac{i}{\hbar}e_*\Phi\right)\Psi = - \cfrac{1}{2m_*} \left(\cfrac{\hbar}{i} \vec{\nabla} - e_*\vec{A} \right)^2 \Psi + \alpha(\vec{r},T) \Psi - \beta(\vec{r},T) \left \vert \Psi \right \vert^2 \Psi \\ \label{GL_SI_1}
\end{equation}
\begin{equation}
\sigma \left(\cfrac{\partial \vec{A}}{\partial t}+\vec{\nabla}\Phi\right) = \cfrac{e_* \hbar}{2m_*i} \left( \Psi^* \vec{\nabla} \Psi - \Psi \vec{\nabla} \Psi^* \right) - \cfrac{e_*^2}{m_*} \left \vert \Psi \right \vert^2 \vec{A} - \cfrac{1}{\mu_0} \vec{\nabla} \times \left ( \vec{\nabla}\times \vec{A} - \vec{B}_a \right ) \\ \label{GL_SI_2}
\end{equation}
\end{widetext}

where $\Psi=\Psi(\vec{r},T,t)$ is the time-dependent order parameter, $\vec{A}=\vec{A}(\vec{r},t)$ is the magnetic vector potential, $\vec{B}_a=\vec{B}_a(\vec{r},t)$ is the externally applied magnetic field, $\Phi=\Phi(\vec{r},t)$ is the scalar electric potential, $D$ is the phenomenological electron diffusion coefficient given by $D=\cfrac{v_Fl}{3}$ \cite{Cyrot1973} with $v_F$ being the Fermi velocity and $l$ being the quasi-particle mean free path \cite{Vinokur2001}, $\sigma$ is the electric conductivity of the normal (non-superconducting) state. It is evident from Eq.(\ref{GL_SI_1}) that $\gamma=\cfrac{\hbar^2}{2m_*D}$ and can also be written as $\gamma=\left \vert \alpha (T) \right \vert \tau_{\Psi}(T)$, where $\tau_{\Psi}(T)=\cfrac{\xi(T)^2}{D} =\cfrac{\pi \hbar}{8 k_B \left ( T_c - T \right )}$ is a characteristic time for the relaxation of the GL order parameter \cite{Hernandez2002} . Here, $\alpha(T)=\alpha(0)\left(1-\cfrac{T}{T_c}\right)$ and $\xi^2=\cfrac{\xi_0^2}{\left(1-\cfrac{T}{T_c}\right)}$, where $\xi_0$ is the zero temperature GL coherence length. \par

Eq.(\ref{GL_SI_1}) was first proposed by Schmid \cite{Schmid1966}, following the derivation of the GL equation from BCS \cite{BCS1957MicTheory, BCS1957Theory} by Gor'kov \cite{Gorkov1968}. Eq.(\ref{GL_SI_2}) is Ampere's law $\vec{\nabla} \times \vec{B}(\vec{r}) = \mu_0 \left( \vec{J_s}(\vec{r}) + \vec{J_n}(\vec{r}) \right)$, where $\vec{J_n}(\vec{r}) = -\sigma \cfrac{\partial \vec{A}(\vec{r})}{\partial t}$ is the normal current and the supercurrent is defined in Eq.\ref{Jscurrent_SI} . \par

The superconducting current can be obtained from the expectation value of the momentum operator for a charged particle in a magnetic field:

\begin{equation}
\vec{J_s}(\vec{r},t)= \cfrac{e_* \hbar}{2m_*i} \left( \Psi^* \vec{\nabla} \Psi - \Psi \vec{\nabla} \Psi^* \right) - \cfrac{e_*^2}{m_*} \left \vert \Psi \right \vert^2 \vec{A} \\ \label{Jscurrent_SI}
\end{equation}

The TDGL equations are invariant under the following change of gauge \cite{Sorensen2011}:
\begin{equation}
 \Psi(\vec{r},t) \rightarrow \Psi(\vec{r},t)e^{i \chi(\vec{r},t)}
\end{equation}
\begin{equation}
 \vec{A}(\vec{r},t) \rightarrow \vec{A}(\vec{r},t)+\cfrac{\hbar}{e_*} \vec{\nabla}\chi(\vec{r},t)
\end{equation}
\begin{equation}
 \Phi(\vec{r},t) \rightarrow \Phi(\vec{r},t)-\cfrac{\hbar}{e_*}\cfrac{\partial\chi(\vec{r},t)}{\partial t} 
\end{equation}

where $\chi(\vec{r},t)$ is any (sufficiently smooth) real-valued scalar function of position and time. One can fix the gauge as $\cfrac{\partial\chi(\vec{r},t)}{\partial t}=\cfrac{e_*}{\hbar} \Phi(\vec{r},t)$ in order to effectively eliminate the electric potential at all times \cite{Sorensen2011, Robson2017Kerr}. \par

It is useful to introduce dimensionless variables (denoted by twiddles $\sim$) to simplify the simulation and normalize Eqs. \ref{GL_SI_1} and \ref{GL_SI_2}:
The order parameter is scaled according to $\Psi_\infty$, $\Psi \rightarrow \Psi_\infty \widetilde{\Psi}$ where $\left \vert\Psi_\infty(\vec{r})\right \vert^2=-\cfrac{\alpha_0}{\beta_0}$ is the bulk superfluid density at zero temperature in the absence of external magnetic field, $\alpha_0 \equiv \alpha (T=0)$ and $\beta_0 \equiv \beta (T=0)$. The spatial coordinates are scaled according to the zero temperature GL penetration depth $\lambda_0 \equiv \sqrt{ \cfrac{m}{\mu_0 n_s {e_*}^2}}$, so that $\left(x,y,z\right) \rightarrow \left( \lambda_0 \widetilde{x}, \lambda_0 \widetilde{y}, \lambda_0 \widetilde{z} \right)$, thus $\vec{\nabla} \rightarrow \cfrac{1}{\lambda_0}\vec{\widetilde{\nabla}}$ \footnote{Note that the zero temperature GL coherence length $\xi_0$ can also be used as a normalization length scale (See Ref. \cite{Vinokur1996,Peng2017})}. Time is scaled according to the characteristic time for the relaxation of the vector potential $\tau_{0}$, $t \rightarrow \tau_{0} \widetilde{t} $ where $\tau_{0} \equiv \mu_0 \lambda_{0}^2 \sigma_n $ \footnote{This time scale is also used in Refs. \cite{Vinokur1996, Hernandez2002, Hernandez2007, Sorensen2011, Blair2018}} and $\sigma_n$ is the normal state conductivity at 0K (as opposed to the conductivity of non-superconducting current at any temperature denoted as $\sigma(T)$). The temperature is scaled according to the critical temperature of the superconductor $T_{c}$, $T \rightarrow T_{c} \widetilde{T}$. The vector potential $\vec{A} \rightarrow \cfrac{\Phi_{0}}{\xi_0} \vec{\widetilde{A}}$ where $\Phi_{0}=\cfrac{h}{2e}$ is the magnetic flux quantum. The superconductor current is scaled in terms of $J_c$, $\vec{J} \rightarrow \cfrac{J_c}{\kappa} \vec{\widetilde{J}}$, where $ J_c= \cfrac{\Phi_{0}}{2 \pi \mu_{0}\lambda_0 \xi_0^2}=\cfrac{ B_{c2}}{\mu_{0} \lambda_0}$ is the critical current density at $T=0$ and $B=0$, and $\kappa$ is the GL parameter and is defined as the ratio of two characteristic length scales $\kappa \equiv \cfrac{\lambda_0}{\xi_0}$. The normal state conductivity is scaled with its zero temperature value $\sigma \rightarrow \sigma_n \widetilde{\sigma}$ and since it is nearly constant in the temperature range of interest for Nb, it is set to $\widetilde{\sigma}=1$. The "normalized friction coefficient" is defined as the ratio between the two characteristic time scales $\tau_{\Psi}$ and $\tau_0$, $\eta \equiv \cfrac{\tau_{\Psi}}{\tau_{0}}$ \cite{Hernandez2002,Blair2018} and is proportional to $\gamma$ defined in Eq.{\ref{gamma}} $\left ( \eta = \cfrac{\gamma}{\vert \alpha \vert \tau_0} \right )$. For cases when the source of externally applied magnetic field is outside of the superconducting domain, the $\vec{B}_a$ term in Eq.(\ref{GL_SI_2}) should be dropped because $\vec{\nabla} \times \vec{B}_a=0$ everywhere within the superconducting domain. \par

Rewriting Eq.(\ref{GL_SI_1}) and Eq.(\ref{GL_SI_2}) using the newly introduced dimensionless quantities and dropping " $\sim$ " we have :

\begin{widetext}
\begin{equation}
\eta \cfrac{\partial \Psi}{\partial t} = - \left(\cfrac{i}{\kappa} \vec{\nabla} + \vec{A} \right)^2 \Psi + \left(\epsilon(\vec{r},T)- \left \vert \Psi \right \vert^2 \right) \Psi \\ \label{GL_norm_1}
\end{equation}

\begin{equation}
\sigma \cfrac{\partial \vec{A}}{\partial t} = \cfrac{1}{2\kappa i} \left( \Psi^* \vec{\nabla} \Psi - \Psi \vec{\nabla} \Psi^* \right) - \left \vert \Psi \right \vert^2 \vec{A} - \vec{\nabla} \times \vec{\nabla} \times \vec{A} \\ \label{GL_norm_2}
\end{equation}

\begin{equation}
\vec{J_s}(\vec{r},t) =\cfrac{1}{2\kappa i} \left( \Psi^* \vec{\nabla} \Psi - \Psi \vec{\nabla} \Psi^* \right) - \left \vert \Psi \right \vert^2 \vec{A} \\
\label{Currentdef}
\end{equation}
\end{widetext}

 Defects (such as pinning sites) can be introduced into the model via spatial variation of the GL coefficient $\alpha(\vec{r},T)$. Such defects could be due to spatial variation of temperature $T$, critical temperature $T_c(\vec{r})$ and/or spatial variation of the mean free path $l(\vec{r})$. One can calculate the vortex pinning potential created by these kinds of disorder using the method outlined in \cite{Vinokur1994}. The pinning coefficient $\epsilon(\vec{r},T) = \cfrac{\alpha(\vec{r},T)}{\alpha(T=0)} = \cfrac{\xi^2(T=0)}{\xi^2(\vec{r},T)} = 1-\cfrac{T}{T_c(\vec{r})}$ dictates the maximum possible value for the superfluid density $n_s(\vec{r},T)$ at a given location and temperature in the absence of external magnetic field. 

In this work we are interested in studying the effects of some common SRF surface defects, such as lossy Nb-oxides and metallic Nb-hydrides near the surface of Nb \cite{Gurevich2017}. These types of defects are either non-superconducting or have lower critical temperature than Nb. Such metallic inclusions, or the effect of nonzero temperature, can be specified through $\epsilon(\vec{r},T)$ \cite{Glatz2015,Milosevich2009, Hernandez2008, Miyamoto2004, Aftalion2001} which can range from $\epsilon(\vec{r},T)=0$ (strong order parameter suppression) to $\epsilon(\vec{r},T)=1$ (full superconductivity). \par

To numerically simulate the superconducting domain, we must specify the boundary conditions for the order parameter, current density, and vector potential. In this work only the superconductor-insulator boundary is considered. Any current passing through the boundary between a superconducting domain and vacuum/insulator would be nonphysical, thus on the boundary $\partial \Omega$ of the superconducting domain $\Omega$ we expect:

\begin{equation}
\vec{J} \cdot \hat{n} = 0\>\>\>\>on\>\>\>\>\partial \Omega\\
\label{Jdotn0}
\end{equation}

Here $\hat{n}$ is unit vector normal to the boundary, and since we expect Eq.(\ref{Currentdef}) to be true even when $\vec{A}=0$ and $\Psi\neq0$ the first boundary condition is \cite{Gennes1999Book, Hernandez2002, Sorensen2011, Blair2018}:

\begin{equation}
\vec{\nabla} \Psi \cdot \hat{n} = 0\>\>\>\>on\>\>\>\>\partial \Omega\\
\end{equation}

Likewise when both $\vec{A}\neq0$ and $\Psi\neq0$, to satisfy Eq.( \ref{Jdotn0}):

\begin{equation}
\left \vert \Psi \right \vert^2 \vec{A} \cdot \hat{n} = 0\>\>\>\>on\>\>\>\>\partial \Omega\\
\end{equation}

leading to

\begin{equation}
\vec{A} \cdot \hat{n} = 0\>\>\>\>on\>\>\>\>\partial \Omega\\
\end{equation}

The third condition generally used is the continuity of magnetic field across an interface.

\begin{equation}
\vec{\nabla} \times \vec{A} = \vec{B}_{external}\>\>\>\>on\>\>\>\>\partial \Omega\\ \label{BC3_SI}
\end{equation}

where $\vec{B}_{external}$ is the externally applied magnetic field.

\section{TDGL in COMSOL} \label{ComsolSec}
COMSOL multiphysics simulation software \cite{Comsol} can be used to solve the TDGL equations in both 2D and 3D domains \cite{Sorensen2011, Peng2017}. The main advantage of COMSOL is the intuitive interface of the software and automatic algorithm optimization. A critical comparison of COMSOL and ANSYS simulation software was previously performed \cite{Salvi2010}, where the authors showed that COMSOL can complete the simulation 10 times faster while reaching similar results. The accuracy of the software has been validated by other researchers as well \cite{Gomes2008, Cardiff2008}. COMSOL has an easy learning curve enabling researchers to use the TDGL model as a tool without spending too much effort on algorithm development \cite{Du1994}. \par

The General Form Partial Differential Equation is one of the equations best suited to be solved by COMSOL multiphysics simulation software and is given as:

\begin{equation}
d \cfrac{\partial \vec{u}}{\partial t} + \vec{\nabla} \cdot \vec{\Gamma} = \vec{F} \label{ComsolPDE}
\end{equation}

Here $\vec{F}$ is the driving term vector, $d$ is the inertia tensor, $\vec{u}$ is a column vector of all unknowns and $\vec{\Gamma}$ is a column vector function of $\vec{u}$. We can rewrite Eq.(\ref{GL_norm_1}) and Eq.(\ref{GL_norm_2}) to be in this form. Redefine $\Psi$ and $\vec{A}$ as:

\begin{equation}
\Psi=v_1+iv_2
\end{equation}

where $v_1$ and $v_2$ are real functions of position and time. 

\begin{equation}
\vec{A}=A_1\hat{x}+A_2\hat{y}+A_3\hat{z}
\end{equation}

where $A_1$, $A_2$ and $A_3$ are real functions of position and time representing the magnitudes of the components of $\vec{A}$ in the $\hat{x}$, $\hat{y}$, $\hat{z}$ directions. \par

We thus have 5 independent unknown variables, and 5 equations (2 from Eq. (\ref{GL_norm_1}), real and imaginary; 3 from Eq. (\ref{GL_norm_2}), 3 vector components ). After some simple mathematical rearrangement we get an equation of the form of Eq.(\ref{ComsolPDE}):

\begin{widetext}
\begin{equation}
\begin{bmatrix} 
\eta&0&0&0&0\\
0&\eta&0&0&0\\
0&0& \sigma&0&0\\
0&0&0&\sigma&0\\
0&0&0&0&\sigma 
\end{bmatrix}
\cdot \frac{\partial}{\partial t}
\begin{bmatrix} 
v_{1}\\v_{2}\\A_1\\A_2\\A_3 
\end{bmatrix}
+\begin{bmatrix} 
\frac{\partial}{\partial x} & \frac{\partial}{\partial y} & \frac{\partial}{\partial x}
\end{bmatrix}
\cdot
\begin{bmatrix} 
-\cfrac{v_{1x}}{\kappa^2} & -\cfrac{v_{1y}}{\kappa^2} & -\cfrac{v_{1z}}{\kappa^2}\\
-\cfrac{v_{2x}}{\kappa^2} & -\cfrac{v_{2y}}{\kappa^2} & -\cfrac{v_{2z}}{\kappa^2}\\
0 & A_{2x}-A_{1y} & A_{3x}-A_{1z} \\
A_{1y}-A_{2x} & 0 & A_{3y}-A_{2y} \\
A_{1z}-A_{3x} & A_{2z}-A_{3y} & 0 
\end{bmatrix}
=\vec{F} \label{ComsolGL}
\end{equation} 

\begin{equation}
\vec{F}=\begin{bmatrix} 
\cfrac{\left( A_{1x}+A_{2y}+A_{3z} \right)}{\kappa} v_2 +\cfrac{2 \left( A_1 v_{2x}+A_2 v_{2y} +A_3 v_{2z} \right)}{\kappa} - (A_1^2+A_2^2+A_3^2) v_1 + \left(\epsilon -\left(v_1^2+v_2^2\right)\right)v_1 \\
-\cfrac{\left( A_{1x}+A_{2y}+A_{3z} \right)}{\kappa} v_1 -\cfrac{2 \left( A_1 v_{1x}+A_2 v_{1y} +A_3 v_{1z} \right)}{\kappa} - (A_1^2+A_2^2+A_3^2) v_2 + \left(\epsilon -\left(v_1^2+v_2^2\right)\right)v_2 \\
\cfrac{\left(v_1v_{2x}-v_2v_{1x}\right)}{\kappa} - \left(v_1^2+v_2^2\right)A_1 \\
\cfrac{\left(v_1v_{2y}-v_2v_{1y}\right)}{\kappa} - \left(v_1^2+v_2^2\right)A_2 \\
\cfrac{\left(v_1v_{2z}-v_2v_{1z}\right)}{\kappa} - \left(v_1^2+v_2^2\right)A_3
\end{bmatrix}
\label{SourceTerm}
\end{equation}
\end{widetext}

Here $v_{1x}$ stands for $\frac{\partial v_1}{\partial x}$, $A_{2z}$ stands for $\frac{\partial A_2}{\partial z}$ and so on. These equations say that the change in $\vec{A}(\vec{r},t)$ is driven by the total current, while the change in $\Psi(\vec{r},t)$ is driven by both $\Psi(\vec{r},t)$ and its interaction with $\vec{A}(\vec{r},t)$. \par

The boundary conditions at the superconductor-vacuum interface are as follows:

\begin{equation}
\vec{\nabla} \Gamma \cdot \hat{n} = 0\>\>\>\>on\>\>\>\>\partial \Omega\\ \label{BC1}
\end{equation}
\begin{equation}
\vec{A} \cdot \hat{n} = 0\>\>\>\>on\>\>\>\>\partial \Omega\\ \label{BC2}
\end{equation}
and
\begin{equation}
\vec{\nabla} \times \vec{A} = \vec{\nabla} \times \vec{A_{ext}}\>\>\>\>on\>\>\>\>\partial \Omega\\ \label{BC3}
\end{equation}

\subsection{Two-Domain TDGL and Inclusion of Superconducting Screening} \label{TwoDomainSec}

\begin{figure}[t]
\centering
\includegraphics[width=0.45\textwidth]{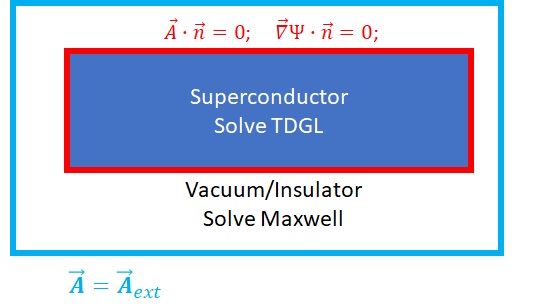}
\caption{Schematic view of the superconductor and vacuum domains and boundary conditions in our TDGL simulations.}
\label{DomainDiagram}
\end{figure}

After reviewing some previously published TDGL simulations \cite{Vinokur1996,Miyamoto2004,Sorensen2011}, we noticed that usually Eq.(\ref{BC3_SI}) or Eq.(\ref{BC3}) are enforced on the boundary of the superconductor. However this implies that the superconducting screening current has no effect on the magnetic field at the boundary and beyond the superconducting domain. This is physically incorrect for the situation of interest to us. The effect of screening currents is crucial when one is trying to simulate spatially nonuniform external magnetic field (like that arising from a nearby magnetic dipole), and the resulting nonlinear response of the superconductor. \par

To include the important physics of screening, our simulation is divided into two domains: Superconductor and Vacuum (Fig. \ref{DomainDiagram}). The full coupled TDGL equations are solved in the superconductor domain, while only Maxwell's equations are solved in the vacuum domain, with appropriate boundary conditions at the interface. Any finite value of $\vec{A} \cdot \vec{n}$ or $\vec{\nabla} \Psi \cdot \vec{n}$ would lead to a finite current passing through the superconductor-vacuum boundary (red box in Fig. \ref{DomainDiagram}), which is nonphysical, hence equations (\ref{BC1}) and (\ref{BC2}) are enforced at the superconductor/vacuum interface. Any externally applied magnetic field is introduced by placing a boundary condition on the outer boundary of vacuum domain (Eq.(\ref{BC3})). The vacuum domain is assumed to be large enough that at the external boundary (blue box in Fig. \ref{DomainDiagram}) the magnetic field generated by the superconductor is negligible. Fig. \ref{DomainDiagram} schematically summarizes this scenario. \par
We now examine several key examples where it is crucial to include the screening response of the superconductor to capture the interesting physics. Through these two examples we validate our approach to solving the TDGL equations.

\subsection{Superconducting sphere in a uniform magnetic field}

\begin{figure}[t]
\centering
\includegraphics[width=0.45\textwidth]{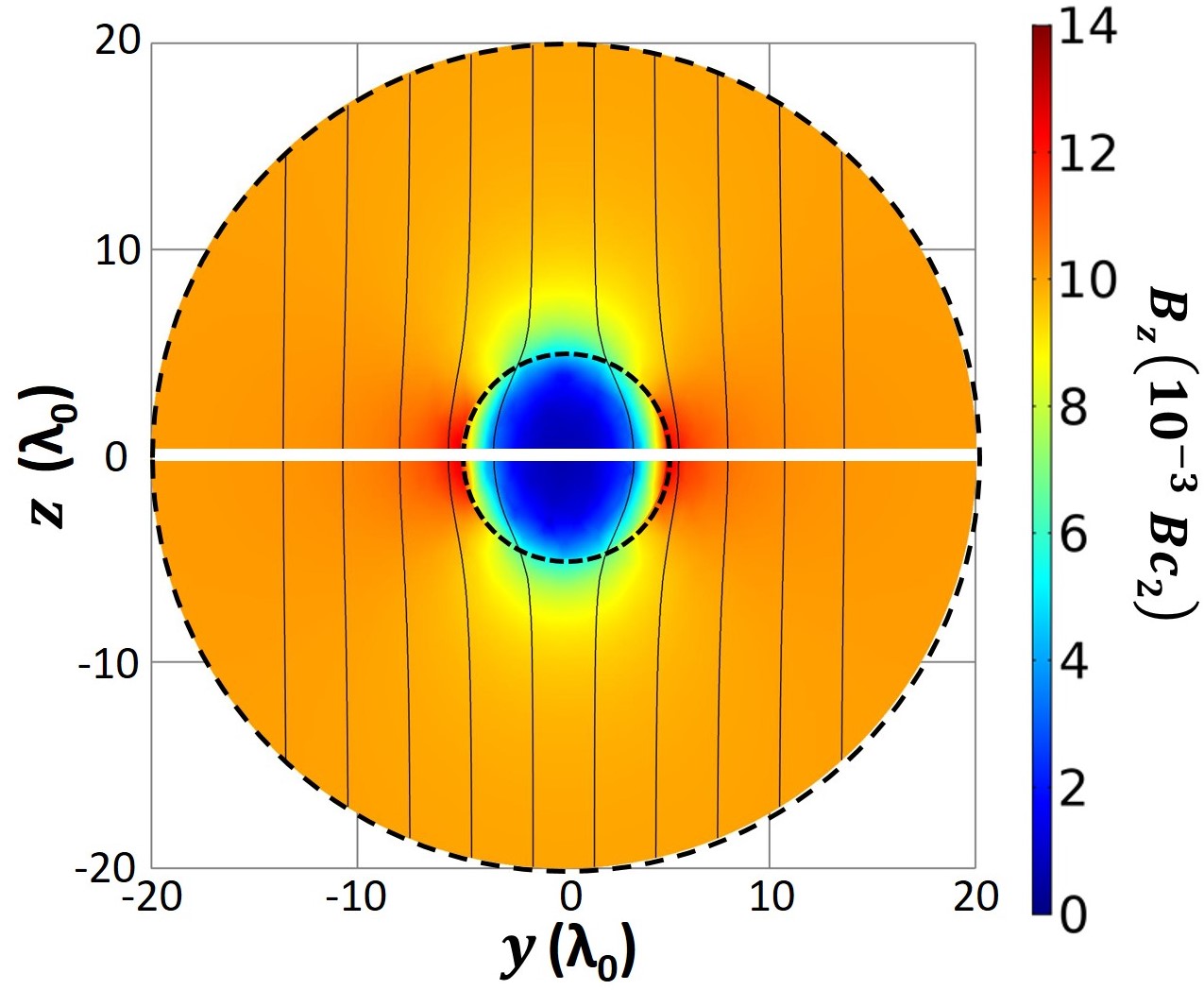}
\caption{Plot of TDGL two-domain solution for z component of magnetic field in a plane through the center of the sphere in and around a superconducting sphere in the Meissner state subjected to a uniform static external magnetic field in the z-direction. The dashed lines show the boundaries of the spheres, with the smaller sphere being the superconducting sphere with diameter $10\lambda_0$ and the larger sphere being the vacuum domain with diameter $40\lambda_0$. The solution is obtained for temperature $T=0$, GL parameter $\kappa=1$ and external magnetic field $\vec{B}_{applied}=10\times10^{-3} B_{c2} \hat{z}$. Black lines show the streamline plot of magnetic field, while the color represents the value of magnetic field component $B_z$. The white line indicates the equator, and the magnetic field along the white line is shown in Fig. \ref{spherefield}.}
\label{sphere}
\end{figure}

First consider the classic problem of a superconducting sphere immersed in a uniform magnetic field. Assume that the superconductor remains in the Meissener state. It is known from the exact solution to this problem that there will be an enhancement of the magnetic field at the equatorial surface of the superconducting sphere due the magnetic flux that is expelled from the interior of the sphere. To test this approach to solving the TDGL equations we created a model of this situation in COMSOL \footnote{For all the simulations presented in this work, the time-dependent study in the COMSOL Multiphysics software was used. The Direct-MUMPS solver with the default parameters was used as the general solver and time-stepping was performed using the Backward Differentiation Formula (BDF) solver. The maximum time-step was constrained to 1. The free tetrahedral mesh was used on the $y>0$ domain, and the same mesh was mirrored in the $y<0$ domain.}. We simulated the response using the two-domain method, and the conventional single domain method used in many other contexts, and then compared both results with the exact analytical solution for the magnetic field profile \cite{Matute1999}. \par

Fig. \ref{sphere} shows the TDGL simulation of a superconducting sphere subjected to a uniform static external magnetic field. The boundaries of the spheres are shown with the dashed lines, where the smaller sphere is the superconducting sphere, and the larger sphere is the vacuum domain. The colors represent the amplitude of the $\hat{z}$-component of the applied magnetic field in the y-z plane passing through the common center of the spheres. Black lines show the streamline plot of magnetic field in the same y-z plane. The streamline plot is defined as collection of lines that are tangent everywhere to the instantaneous vector field, in this case to the direction of the magnetic field. The simulation was initialized in a field free configuration and the external magnetic field was applied at $t=0$. The simulation was iterated for $t=1000\tau_0$ time steps after which the changes in $\left \vert\Psi\right \vert^2$ were $<0.1\%$ per iteration. \par

To test the reproducibility of the result, the simulation was later repeated, but this time the external magnetic field was increased linearly in time from zero to $0.01 B_{c2}$ between time $0$ and $500 \tau_0$. After this, the simulation was again iterated for $t=1000\tau_0$ time steps. The results of these two simulations were identical. \par

Eqs. (\ref{BC1}) and (\ref{BC2}) were enforced on the spherical superconductor-vacuum boundary ($r=5\lambda_0$) in both cases. When the two-domain method was used, the TDGL equations were solved in the inner sphere ($r<5\lambda_0$) and only Maxwell's equations were solved in the vacuum domain ($5\lambda_0<r<20\lambda_0$ ). Eq.(\ref{BC3}) was enforced at the outer boundary of the simulation ($r=20\lambda_0$). When the single domain simulation method was used, Eq.(\ref{BC3}) was enforced at the inner boundary of the simulation ($r=5\lambda_0$) and the vacuum domain was not utilized. \par

\begin{figure}[t]
\centering
\includegraphics[width=0.5\textwidth]{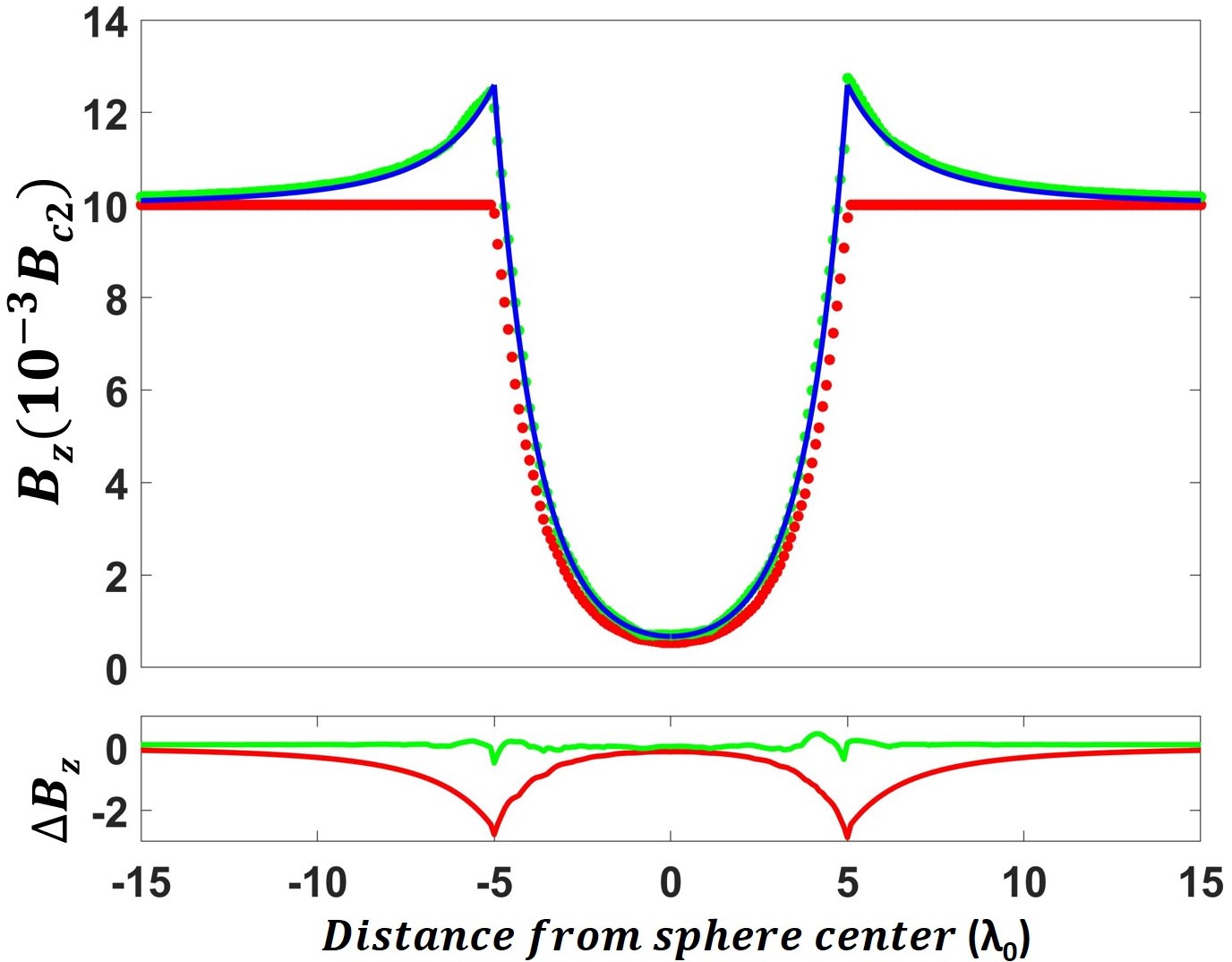}
\caption{Top: Magnetic field $\hat{z}$-component ($B_z$) profile through the center of sphere (white line in Fig \ref{sphere}). The results of a single domain TDGL model are shown in red, a two-domain TDGL model in green, and the analytic solution is shown as a blue solid line. Bottom: The difference between a two-domain TDGL model and the analytic solution is shown in green and the difference between single domain TDGL model and the analytic solution is shown in red. The biggest difference is observed at the surface.}
\label{spherefield}
\end{figure}

The top plot in Fig. \ref{spherefield} shows the profile of the z-component of magnetic field ($B_z$) along a line through the center of the sphere, in a plane perpendicular to the externally applied magnetic field (white line in Fig. \ref{sphere}) calculated from the single domain simulation, the two-domain simulation and the analytic result. Inside the sphere, the magnetic field profile calculated from the single domain simulation and the two-domain simulation are very similar although not identical. The bottom plot in Fig. \ref{spherefield} shows the difference between the TDGL simulation results and the analytic solution. The field deep inside the sphere is strongly suppressed by the screening currents. This can also be seen from the color-map in Fig. \ref{sphere}. The blue region inside the sphere corresponds to the fully shielded portion of the sphere. However, there is a region outside the sphere around the equator where the magnetic field is enhanced (red color in Fig. \ref{sphere}). \par

At the surface of the sphere the magnetic field calculated from the two-domain model reproduces the exact analytic solution, while the single domain model fails to account for the enhancement of magnetic field on the equator of the sphere. This disparity between the single domain model and analytic solution is caused by the treatment of the boundary conditions. In the single domain model, Eq. \ref{BC3} is enforced at the superconductor-vacuum interface, which completely ignores the effect of screening currents. Thus a two-domain model should be used for any problem where screening and the magnetic field profile at the surface of the superconductor is important. \par

\subsection{Point magnetic dipole above a semi-infinite superconductor \label{PointDipoleSec}}

To ensure that we can accurately simulate the screening currents produced by a spatially nonuniform magnetic field, we numerically simulated the case of a static point magnetic dipole placed at a height of $h_{dp}=1\lambda_0$ above the surface of a semi-infinite superconductor. The superconducting domain and vacuum domain are simulated inside two coaxial cylinders with equal radius $R=8\lambda_0$ with common axis along the $\hat{z}$ direction of the Cartesian coordinate system. The origin of this coordinate system is located on the superconductor surface immediately below the dipole. The thickness of the superconducting domain is $h_{sc}=10 \lambda_0$ and the height of the vacuum domain is $h_{vac}=5\lambda_0$. The normalized friction coefficient $\eta$ and the GL parameter $\kappa$ are set to $1$. \par

The surface magnetic fields produced by the dipole are assumed to be below the lower critical field $H_{c1}$, so that the superconductor remains in the Meissner state. The simulation was started with a superconductor in the uniform Meissner state and the dipole field equal to 0. Then, at time $t=0$, the dipole magnetic field is turned on, and the simulation is iterated in time until the relative tolerance of $\cfrac{\partial u}{u} < 0.001$ is achieved for all the variables in the column vector of all unknowns $u$ (Eq. \ref{ComsolPDE}). At this point the static solution to the problem is obtained. Later the simulation was repeated with external magnetic field linearly increasing with time over $t=0-500\tau_0$ time interval before reaching a set constant value. The results of these two simulations were identical. \par

\begin{figure}[t]
\centering
\includegraphics[width=0.5\textwidth]{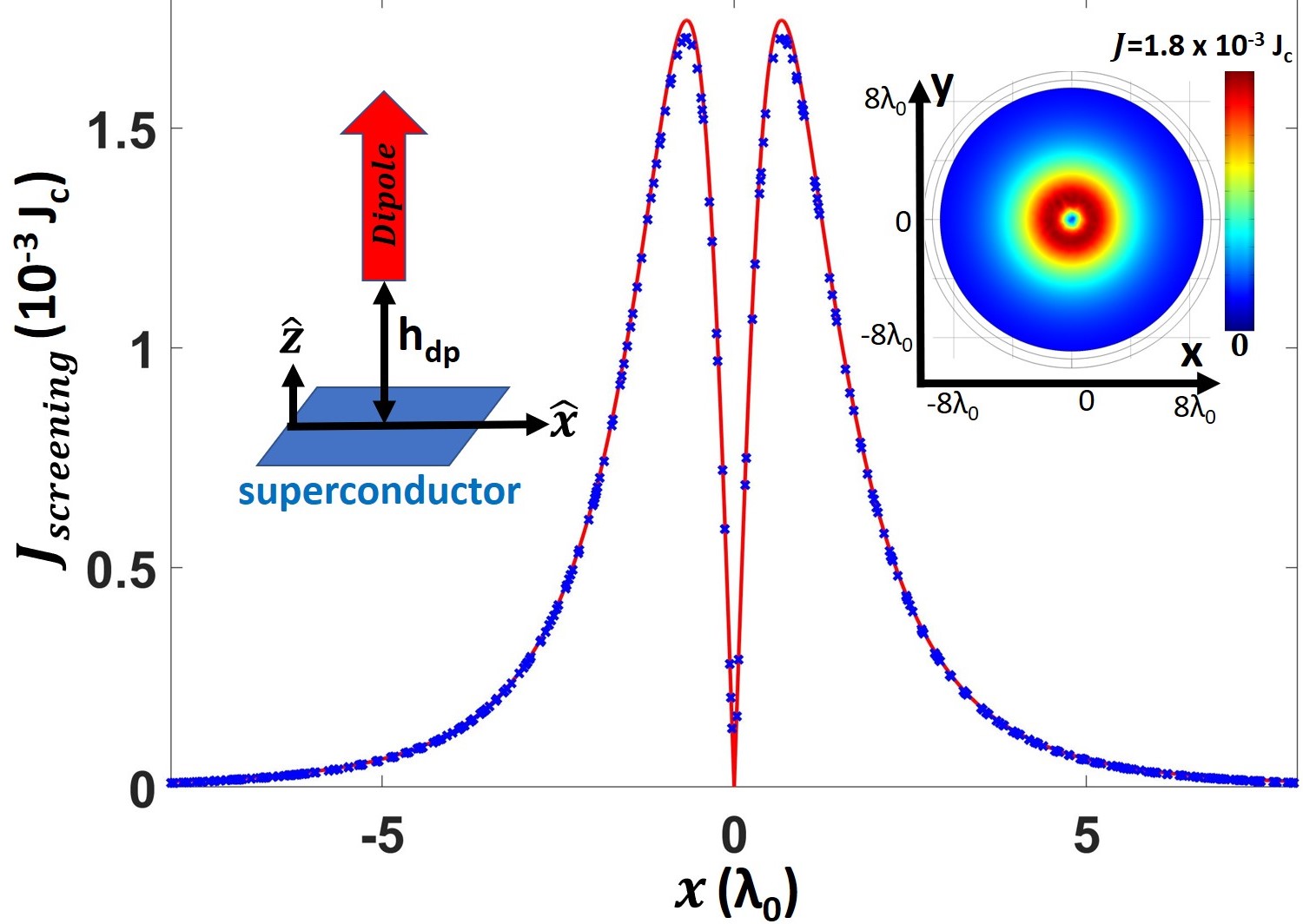}
\caption{ The magnitude of the superconducting screening current density at the surface $J_{screening}$ as a result of a perpendicular magnetic dipole placed $h_{dp}=1 \lambda_0$ above the superconductor vs the horizontal distance from the dipole location obtained from TDGL simulation (blue $\times$) and numerical solution for the same scenario obtained from Ref. \cite{Melnikov1998} (red solid line). Left inset shows a schematic of the dipole over the superconductor, while the right inset shows the top view of the surface current distribution calculated by TDGL, which is azimuthally symmetric. The parameters of the simulation are listed in Table I.}
\label{DipoleCompare}
\end{figure}

We compared our TDGL results for the distribution of the surface screening current density $\vec{J}_{screening}(x,y)$ to numerical results obtained by Melnikov \cite{Melnikov1998} for the case of a perpendicular magnetic dipole. Fig. \ref{DipoleCompare} shows a comparison of the calculated screening current profiles. Both results show that there is a circulating screening current centered directly below the dipole. Also note that the screening current reaches zero at the outer boundary of the simulation. This indicates that a sufficiently large domain was chosen for simulation and no finite size effects are expected. We have very good agreement between the two-domain TDGL simulation result and numerical results obtained by Melnikov, in the low magnetic field limit where there are no vortices (Fig. \ref{DipoleCompare}). This, and the previous result, serve to validate our two-domain approach to properly capturing the screening response of the superconductor in TDGL. \par

\section{Application: Nonlinear Near-Field Magnetic Microwave Microscopy of a Superconductor} \label{ApplicationSec}

The dominant material used in SRF cavities is Nb, which is a type II superconductor and can host vortices. Vortices can be created by high rf magnetic fields used in SRF cavity operation and point-like surface defects \cite{Cooley2011, Iwashita2008}. Vortices can also form due to flux trapped during the cool down procedure. Recent studies showed that the trapped magnetic flux amount depends on the rate at which the cavity is cooled down through the critical temperature and the level of the ambient magnetic field \cite{Romanenko2014}. Decreasing the trapped magnetic flux amount leads to better cavity performance. \par
The type of vortices inside an SRF cavity and the dynamics of those vortices was theoretically studied by Gurevich and Ciovati \cite{Gurevich2008}. For large parallel surface rf magnetic fields and a point-like surface defect, a vortex first enters the superconductor as a vortex semiloop. To study the dynamics of these vortex-semiloops a novel near-field magnetic microwave microscope was successfully built using a magnetic writer from a conventional magnetic recording hard-disk drive \cite{Lee2005GB, Mircea2009, Tai2011, Tai2013, Tai2014, Tai2015, Oripov2019}. A magnetic write head can produce $B_{RF}\approx600mT$ rf magnetic field localized to a $\approx100nm$ length scale \cite{SeagateInfo}. In the experiment, a Seagate perpendicular magnetic writer head is attached to a cryogenic XYZ positioner and used in a scanning probe fashion. Probe characterization results and other details can be found in \cite{Tai2011, Tai2013, Tai2014, Tai2015, Oripov2019}. The probe produces an rf magnetic field perpendicular to the sample surface. The sample is in the superconducting state, so to maintain the Meissner state a screening current is induced on the surface. This current generates a response magnetic field which is coupled back to the same probe, creates a propagating signal on the attached transmission line structure, and is measured with a spectrum analyzer at room temperature. Since superconductors are intrinsically nonlinear \cite{Sauls1995}, both linear and nonlinear responses to an applied rf magnetic field are expected. In said experiment, mainly the third-harmonic response to the inhomogeneous driving field is measured. \par

The rf magnetic field produced by the magnetic writer probe sitting on top of a sample is very similar to the magnetic field produced by a horizontal point magnetic dipole with normalized magnetic moment $M_{dp}(t) || \hat{x}$ placed at a height $h_{dp}$ above the sample. The normalized vector potential produced by such a dipole in free space is given by \cite{Chow2006Book}:

\begin{equation}
\vec{A}_{dp}(x,y,z,t)=\cfrac{M_{dp}(t)}{\big(x^2+y^2+(z-h_{dp})^2\big)^{3/2}}\Big(-(z-h_{dp})\hat{y}+y\hat{z}\Big) \\
\label{DipoleField}
\end{equation}

where the origin of the coordinate system is on the superconductor surface immediately below the dipole. While this is very different from a uniform and parallel magnetic field inside an actual SRF cavity, the dynamics of the vortex semiloops created by this field should be very similar. \par

\begin{figure}[t]
 \centering
 \includegraphics[width=0.45\textwidth]{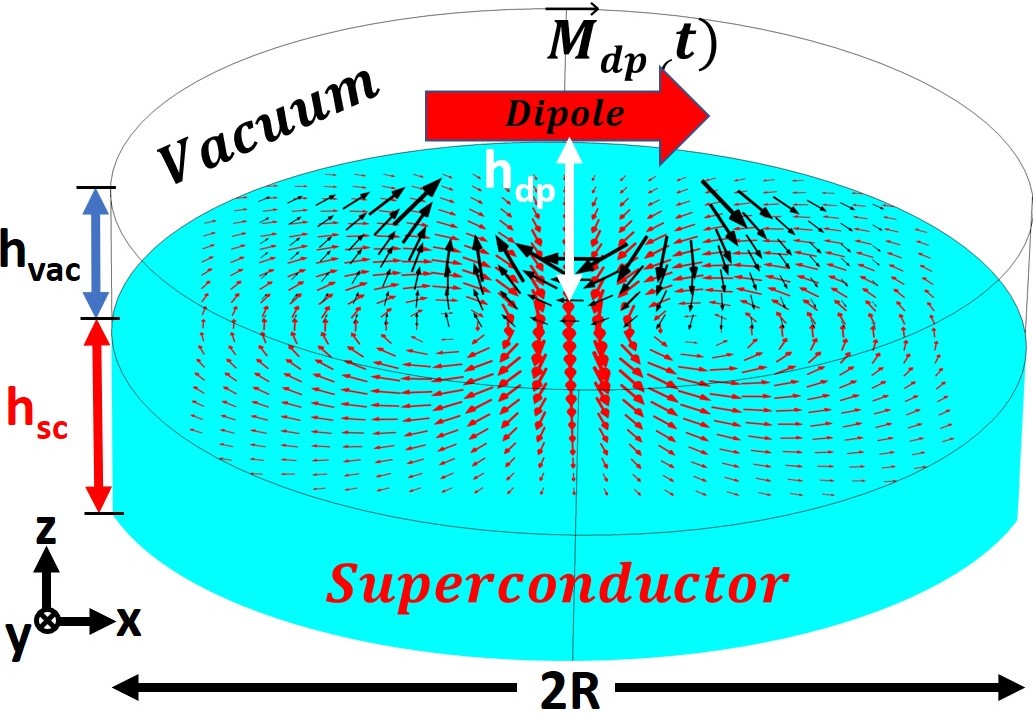}
 \caption{TDGL simulation setup for an oscillating horizontal magnetic dipole $\vec{M}_{dp}$ at height $h_{dp}$ above the superconductor surface. The magnetic probe is approximated as an oscillating point magnetic dipole parallel to the surface. Red Arrows: Surface currents on the horizontal (xy) superconductor/vacuum interface as calculated from the self-consistent TDGL equations. Black Arrows: Externally applied magnetic field on a vertical plane (xz) perpendicular to the superconductor surface and including the dipole.}
 \label{ComsolSetup}
\end{figure}

The superconducting domain and vacuum domain are simulated inside two coaxial cylinders with equal radius $R$ (see Fig. \ref{ComsolSetup}) with common axis along the $\hat{z}$ direction of the Cartesian coordinate system. The thickness of the superconducting domain is $h_{sc}$ and the height of the vacuum domain is $h_{vac}$ in normalized units. \par

The boundary condition Eq. \ref{BC3} is enforced at the top of the vacuum domain, whereas a $\vec{B}=0$ boundary condition is enforced at the bottom and the sides of the superconducting domain, since it is expected that the superconducting currents due to the Meissner state will fully shield the externally applied magnetic field before it reaches the outer boundary of the superconductor. \par
 
The interaction between the probe and the sample was modeled by solving the TDGL equations. In the simulation, we specify $M_{dp}(t)$ indirectly through the the magnetic field experienced at the origin (on the superconductor surface immediately below the dipole) $\vec{B}_{0}(t)=\vec{\nabla} \times \vec{A}_{dp}(0,0,0,t)= - \cfrac{M_{dp}(t) }{h_{dp}^3} \hat{x}$, where $M_{dp}(t)=M_{dp}(0)sin(\omega t)$. The driving rf magnetic field profile is specified through the analytic equation for the magnetic vector potential of a point dipole (Eq. \ref{DipoleField}), therefore the dipole itself can be placed either inside the vacuum domain $h_{vac}>h_{dp}$ or beyond it $h_{vac}<h_{dp}$ without affecting the accuracy of the simulation. $h_{vac}$ is chosen to be large enough to be consistent with Eq. \ref{BC3} at the top of the vacuum domain. \par
The main objective of this work is to simulate the response of the SRF grade Nb, thus the parameters are chosen accordingly. For Nb, $\sigma_n$ ranges from $2\times10^8S/m$ to $2\times10^9S/m$ depending on the RRR value of the material, and $\lambda_0 = 40nm$ \cite{Transtrum2017, Roy2018}. The characteristic time for the relaxation of $\vec{A}$, $\tau_{0} = \mu_0 \lambda_{0}^2 \sigma_n = 4\times 10^{-12} s$ for Nb bulk samples in the clean limit $(RRR\approx300)$. Consequently, the 100-2000 $\tau_0$ range for the period of the magnetic dipole corresponds to a frequency range of $125MHz-2.5GHz$. Hence, the period of the dipole rf magnetic field was chosen to be $\cfrac{2\pi}{\omega}=200 \tau_0$. The GL parameter $\kappa=1$ \cite{Singer2015}, and $\eta$ is on the order of unity (Parameters summarize in Table.I). It should be noted that the relaxation time $\tau_0 \sim ps$ with $\eta=\cfrac{\tau_{\psi}}{\tau_0}\sim 1$ is "fast" in the sense that the order parameter will quickly follow any variations in rf field or current. \par

The spatial distribution of the magnetic field at the surface of the superconductor is set through the value for the dipole height $h_{dp}$. While the driving rf magnetic field is specified through the analytic equation Eq. \ref{DipoleField}, the goal is to reproduce the actual spatial distribution produced by the magnetic writer head at the surface of the superconductor, which was provided by the manufacturer \cite{SeagateInfo}. To produce similar spatial distribution of the magnetic field, we set the dipole height to the $300nm-500nm$ range which corresponds to $h_{dp}$ of $8-12 \lambda_0$ in normalized units. \par

\subsection{The evolution of vortex semiloops with time}

\begin{figure}[t]
\centering
\includegraphics[width=0.5\textwidth]{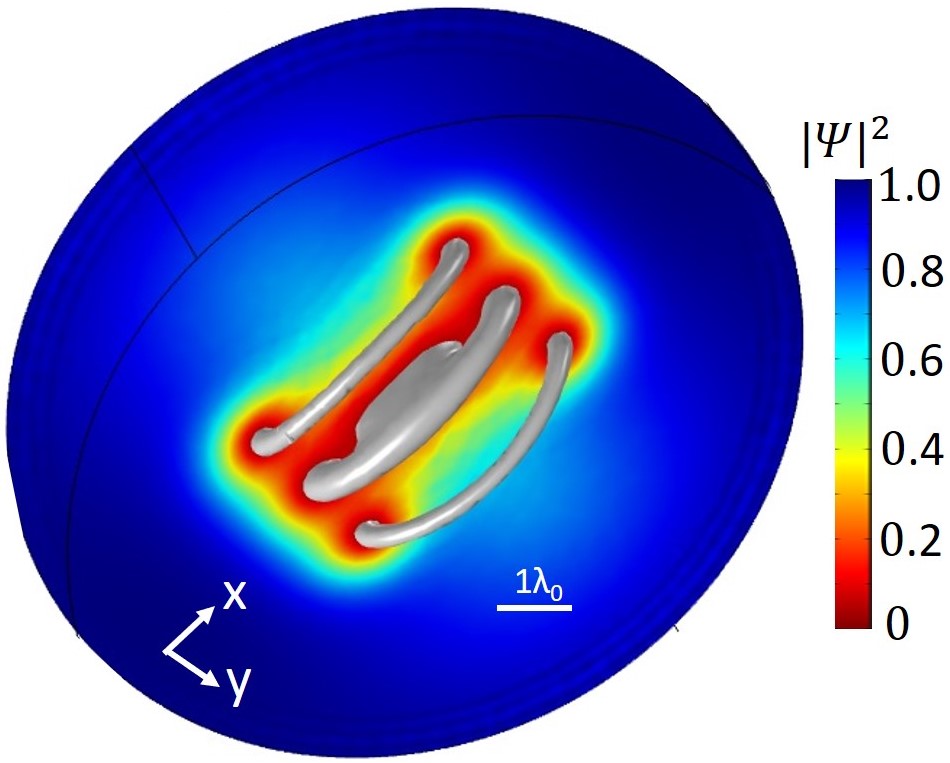}
\caption{Snapshot of 3 vortex semiloops at time $t=73\tau_0$ during the rf cycle of period $200 \tau_0$. In this view, one is looking from inside the superconducting domain into the vacuum domain. Plots of $\left \vert \Psi \right \vert^2$ are evaluated at the superconductor surface for an oscillating parallel magnetic dipole above the superconductor. The three-dimensional silver surfaces (corresponding to $\left \vert \Psi \right \vert^2 =0.005$) show the emergence of vortex semiloops. The simulation parameters are given in Table I .}
\label{Vortex3D}
\end{figure}

We consider a dipole that oscillates sinusoidally in time with frequency $\omega$, and calculate the response of the superconductor to this external inhomogeneous and time-dependent magnetic field. Our objective is to describe a spatially-inhomogenious microwave frequency stimulus of the superconducting surface. In this section a uniform superconductor domain with no defects is considered. The simulation is started with the order parameter having a uniform value of $\vert \Psi \vert^2 = \epsilon(T)$ everywhere. At time $t=0$ the externally applied magnetic field is turned on. Then the simulation is run for several rf cycles to reach the steady state solution. \par

Fig. \ref{Vortex3D} shows the results for such a simulation, and the parameters are given in Table I . The simulation was run for 3 driving periods to stabilize and the results shown in Fig. \ref{Vortex3D} are from the $4^{th}$ driving period. Three well-defined vortex semiloops are illustrated by the three-dimensional silver surface corresponding to $\left \vert \Psi \right \vert^2 =0.005$. \par

\begin{figure}[t]
\centering
\includegraphics[width=0.5\textwidth]{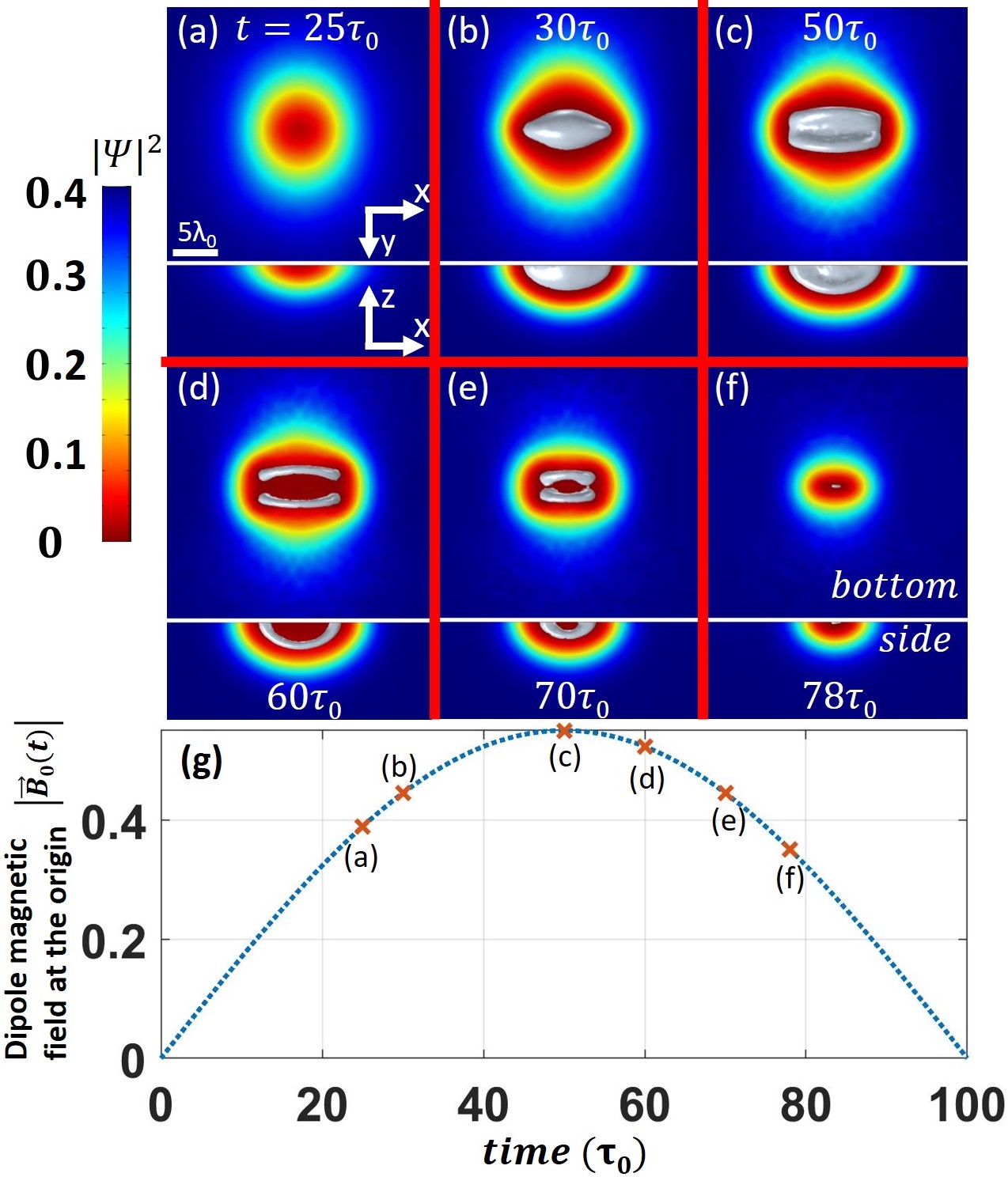}
\caption{ Summary of TDGL solution for an oscillating parallel magnetic dipole above a superconducting surface. (a)-(f) Plots of $\left \vert \Psi \right \vert^2$ evaluated at the superconductor surface at different times for an oscillating parallel magnetic dipole above the superconductor. In the top part of each panel, one is looking from inside the superconducting domain into the vacuum domain, whereas in the bottom part of each panel, one is looking at the x-z cross-section plane towards the $+y$ axis. $\vec{M_{dp}}(t)$ is chosen such that $\vec{B_{0}}(t)=0.55 sin(\omega t) \hat{x}$. The three-dimensional silver surfaces (corresponding to $\left \vert \Psi \right \vert^2 =0.005$) show the emergence of vortex semiloops. (g) $\left \vert \vec{B_{0}} \right \vert$ at the surface vs time during the first half of the rf cycle. Red crosses correspond to field values for snapshots (a)-(f).}
\label{DipolevsTime}
\end{figure}

Fig. \ref{DipolevsTime} shows results for a similar simulation, illustrating the order parameter space and time dependence, and the parameters are given in Table I. The simulation was run for 5 driving periods to stabilize, and the results shown in Fig. \ref{DipolevsTime} are from the $6^{th}$ driving period. We see that as $\vec{B}_{0}(t)$ increases a suppressed $\left \vert \Psi \right \vert^2$ domain (red region) forms at the superconductor surface immediately below the dipole. At $t=50\tau_0$ the magnetic field reaches its peak value and the suppressed superconducting region reaches its deepest point inside the superconducting domain illustrated by the silver surface in Fig. \ref{DipolevsTime}(c). Later ($t>50\tau_0$), the amplitude of the external driving magnetic field decreases, the suppressed $\left \vert \Psi \right \vert^2$ domain rapidly diminishes and vortex semiloops spontaneously emerge, become well-defined (Fig. \ref{DipolevsTime}.d,e), then move back towards the surface and vanish there before the end of the first half of the rf cycle. In the second part of the rf cycle, the same process is repeated but now antivortex-semiloops enter the superconducting domain. The full solution animated over time is available online \cite{VortexSemiloopTimeAnimation}. In this particular scenario vortices and anti-vortices never meet, unlike the situation discussed in \cite{Tai2013Thesis}. \par

\begin{figure}[t]
\centering
\includegraphics[width=0.5\textwidth]{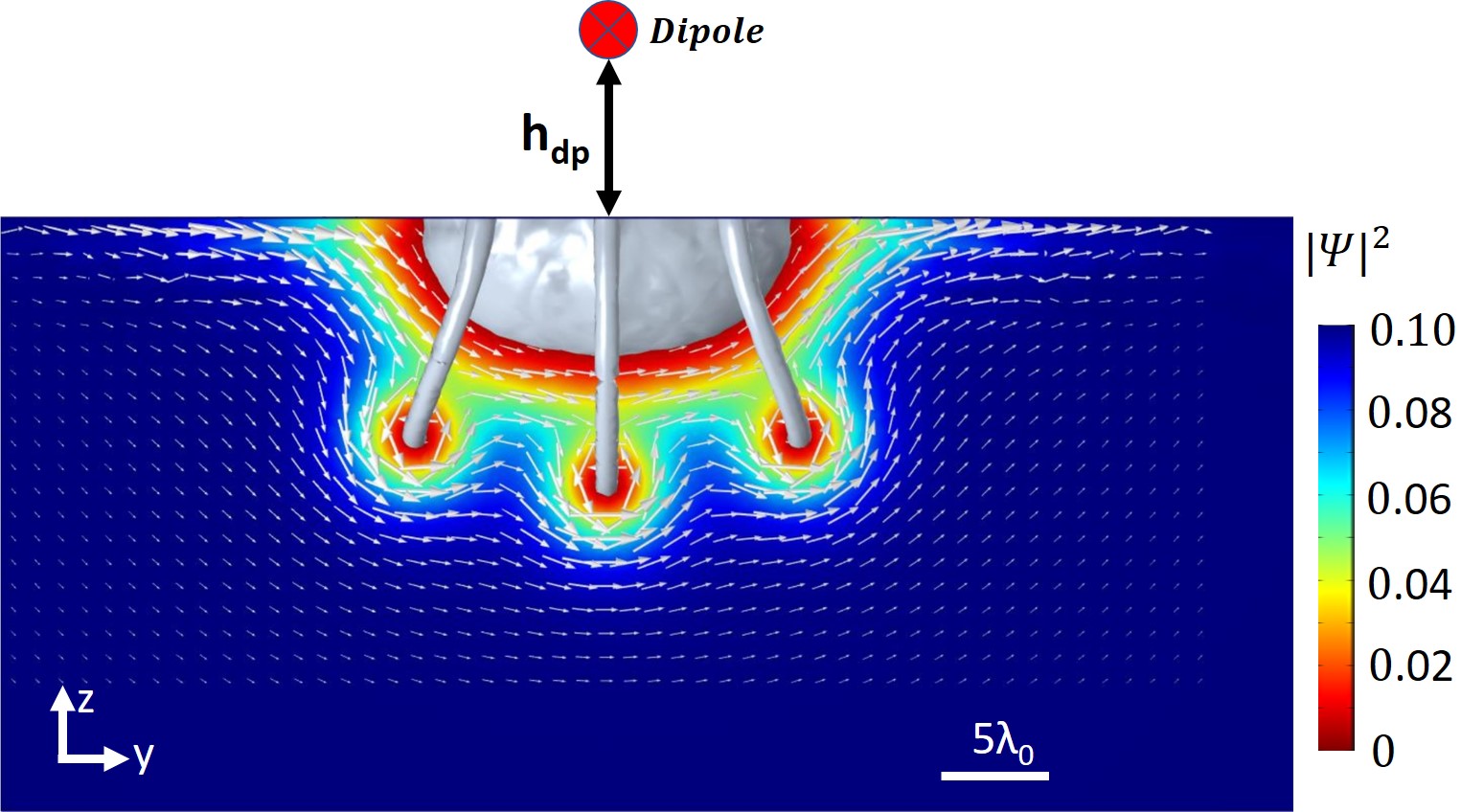}
\caption{Plots of $\left \vert \Psi \right \vert^2$ (color) and $\vec{J}_{surf}$ (arrows) evaluated at the two dimensional $x=0$ plane inside the superconductor at $t=50\tau_0$ for an oscillating parallel magnetic dipole above the superconductor. White arrows indicate the currents induced inside the superconducting domain. The three-dimensional silver surfaces (corresponding to $\left \vert \Psi \right \vert^2 =0.005$) show the emergence of vortex semiloops and the suppressed superconducting domain. All model parameters are listed in Table I.}
\label{DipoleExample_J_E}
\end{figure}

Fig. \ref{DipoleExample_J_E} shows another simulation result with a different set of parameters (listed in Table I). Here the dipole is further away from the surface, at $h_{dp}=12$ and the temperature is set to $T=0.9 T_c$. Three dimensional silver contour surfaces correspond to $\left \vert \Psi \right \vert^2 =0.005$. The two-dimensional screening currents (white arrows) and two-dimensional order parameter (colors) are plotted in the yz-plane. Three vortex semiloops are clearly visible in this $x=0$ cross-section cut. We see that the vortex semiloops penetrated somewhat deeper into the superconductor than the suppressed order parameter domain. 

\subsection{The evolution of vortex semiloops with rf field amplitude}

\begin{figure}[t]
\centering
\includegraphics[width=0.5\textwidth]{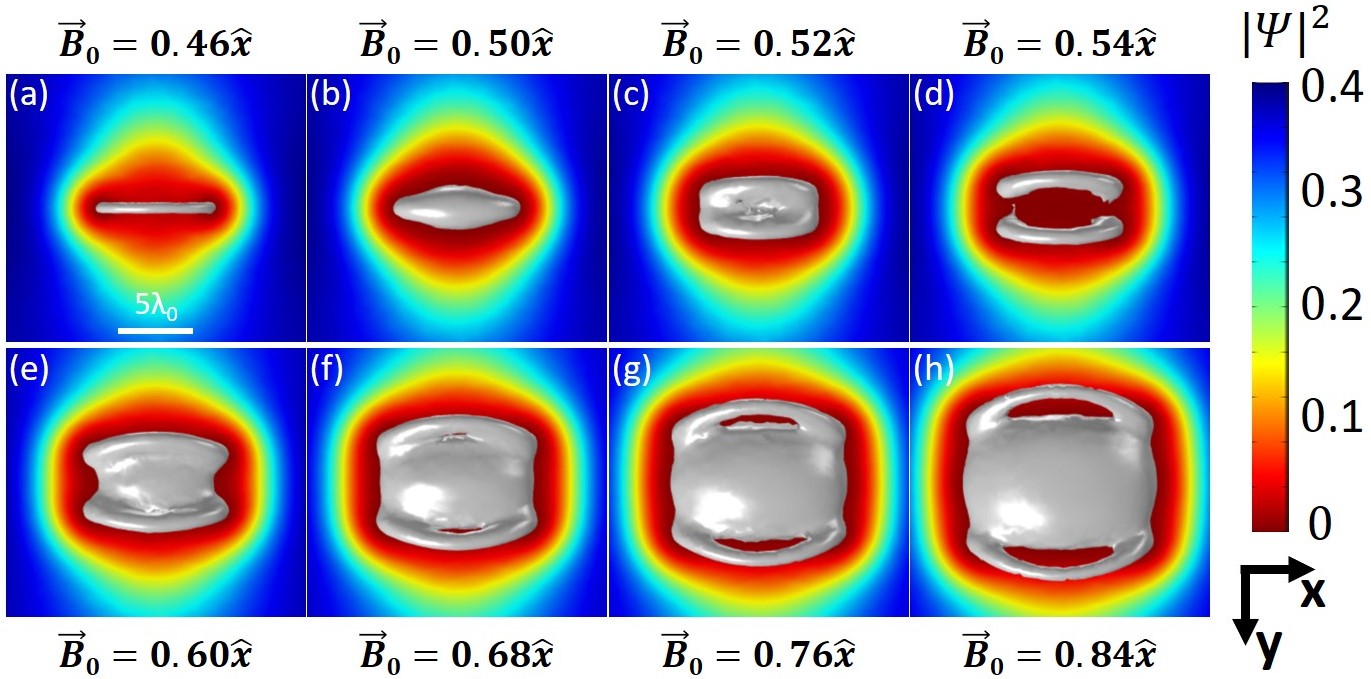}
\caption{a-h) Plots of $\left \vert \Psi \right \vert^2$ evaluated at the superconductor surface at $t=50 \tau_0$ for an oscillating parallel magnetic dipole above the superconductor as a function of dipole strength. In this view, one is looking from inside the superconducting domain into the vacuum domain. The maximum amplitude of applied rf field is shown as $\left \vert \vec{B}_0 \right \vert$. The silver three-dimensional surfaces correspond to $\left \vert \Psi \right \vert^2=0.005$ and show the suppressed order parameter domain and the vortex semiloops. The parameters of the simulation are listed in Table I.}
\label{DipolevsField}
\end{figure}

One can also study the effect of the applied rf field amplitude, defined through $ \left \vert \vec{B}_{0} \right \vert$, on the number and the dynamics of vortex semiloops. Fig. \ref{DipolevsField} shows the bottom view of the order parameter on the surface of the superconducting domain for different values of the applied rf magnetic field amplitude, all at the same point in the rf cycle ($t=50 \tau_0$ and $\vec{B}_0(t)$ at its peak value). As expected, the number of vortex semiloops increase with increasing $ \left \vert \vec{B}_{0} \right \vert$. Once $ \left \vert \vec{B}_{0} \right \vert=0.6$ is reached, a normal state $\left \vert \Psi \right \vert^2=0$ domain emerges at the origin, as opposed to a suppressed $\left \vert \Psi \right \vert^2$ domain observed at lower rf field amplitudes. The full solution as a function of peak applied magnetic field amplitude is available online \cite{VortexSemiloopFieldAnimation}.

\subsection{The effect of localized defects on rf vortex semiloops}
\begin{figure}[t]
\centering
\includegraphics[width=0.5\textwidth]{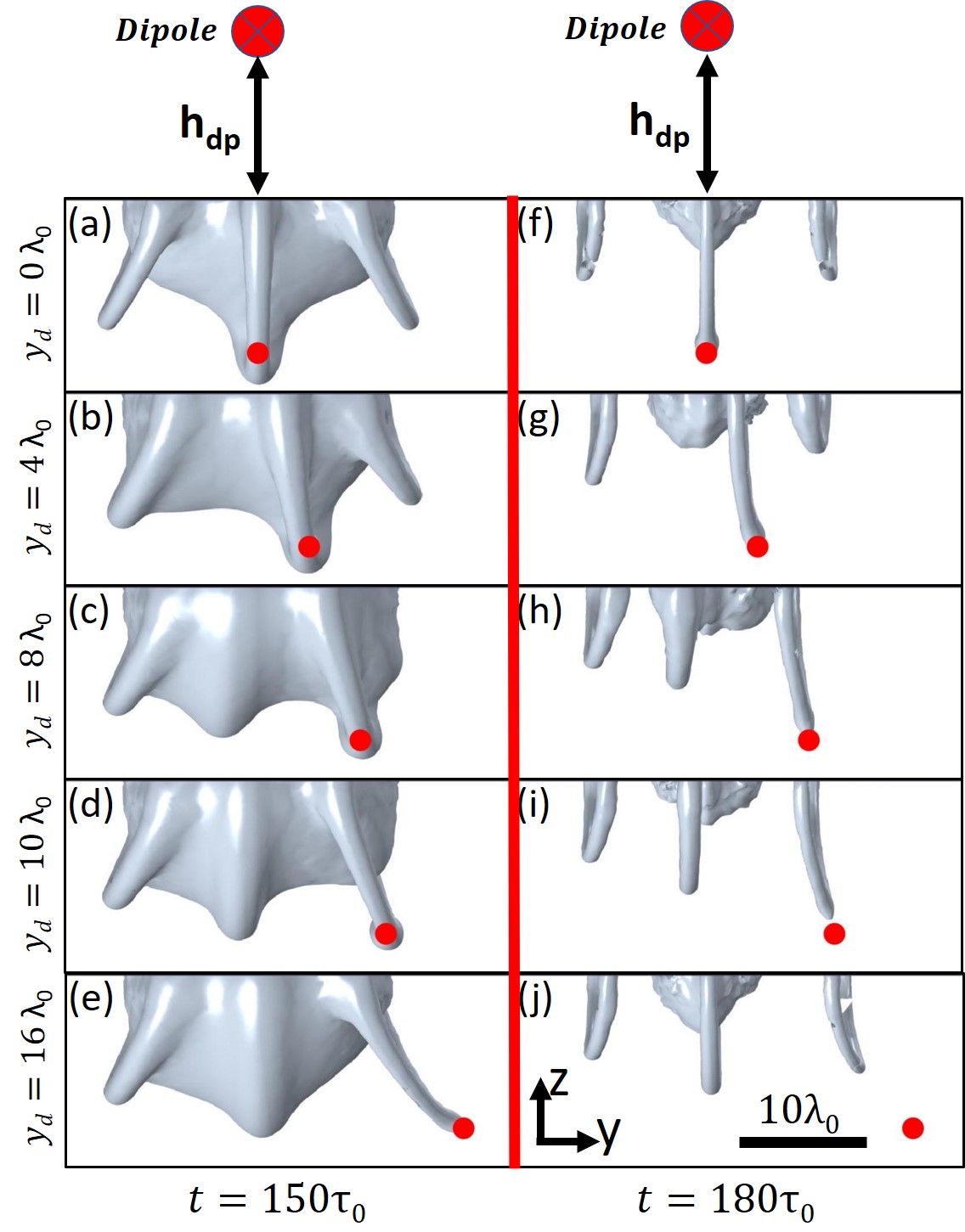}
\caption{Summary of TDGL solutions for an oscillating parallel magnetic dipole above a superconducting surface in the presence of a localized defect at $\vec{r}_d=0\hat{x}+y_d \hat{y}-12\hat{z}$, where $y_d$ is varied from $0$ to $16 \lambda_0$. (a)-(e) Plots of vortex semiloops in the y-z cross-section plane below the dipole illustrated with a three-dimensional silver surface (corresponding to $\left \vert \Psi \right \vert^2 =0.003$) at time $t=150\tau_0$, when the applied magnetic field reaches its peak amplitude. (f)-(j) Plots of vortex semiloops at time $t=180\tau_0$. The defect is denoted by the red dot to the right of the center. $\vec{M_{dp}}(t)$ is chosen such that $\vec{B_{0}}(t)=0.30 sin(\omega t) \hat{x}$. The full list of simulation parameters is given in Table I.}
\label{Defect_Evol}
\end{figure}

 In the past GL has been used to estimate the surface superheating field of superconductors \cite{Transtrum2011} and TDGL was used to study RF vortex nucleation in mesoscopic superconductors \cite{Hernandez2008}. Here we wish to examine the effect of a single point-like defect on rf vortex nucleation in a bulk sample. \par

The effect of a localized defect can be specified through the function $\epsilon(\vec{r},T)=\cfrac{\alpha(\vec{r},T)}{\alpha (T=0)}$ in Eqs. (\ref{GL_norm_1}) and (\ref{SourceTerm}), which can range from $\epsilon(\vec{r},T)=0$ (strong suppression of superconductivity) to $\epsilon(\vec{r},T)=1$ (fully superconducting). Here, $\alpha(\vec{r},T)$ dictates the maximum possible value for the superfluid density $n_s(\vec{r},T)$ in the absence of external magnetic field. A simple defect can be created, for example by defining a Gaussian-in-space domain with suppressed superconducting critical temperature $T_{cd}$, where $0<T_{cd}<1$:

\begin{equation}
\epsilon(\vec{r},T)=1-\cfrac{T}{1-\left( 1-T_{cd} \right) e^{-\frac{(x-x_d)^2}{2\sigma_x}-\frac{(y-y_d)^2}{2\sigma_y}-\frac{(z-z_d)^2}{2\sigma_z}}}.
\label{DefectEq}
\end{equation}

where $( x_d,y_d,z_d )$ are the central coordinates of the defect and $\sigma_x, \sigma_y, \sigma_z$ are the standard deviations in the 3 coordinate directions, all expressed in normalized values. Fig. \ref{Defect_Evol} shows a simulation which was done with parameters given in Table I . A localized defect with $\sigma_x=\sigma_y=\sigma_z=\sqrt{2}$ and $T_{cd}=0.2$ is located at $\vec{r}_d=0\hat{x}+y_d \hat{y}-12\hat{z}$, where $y_d$ is varied from 0 to $16 \lambda_0$, to represent a localized defect that is centered 12 penetration depths ($\lambda_0$) below the surface and offset various distances from the oscillating dipole. We observed very similar vortex semiloops in the time domain evolution as those shown above. However, one of the vortex semiloops is now attracted towards the defect location (shown as red dot in Fig.\ref{Defect_Evol}) and is distorted in shape. Furthermore, the vortex attracted by the defect remains inside the superconductor longer compared to the other vortex semiloops. Note that the semiloop disappears at the end of each half of the rf cycle, hence the pinning potential of this defect is not strong enough to trap the vortex semiloop, only to modify the rf behaviour.

\subsection{Surface Defect in a Parallel rf magnetic field}

\begin{figure}[t]
\centering
\includegraphics[width=0.5\textwidth]{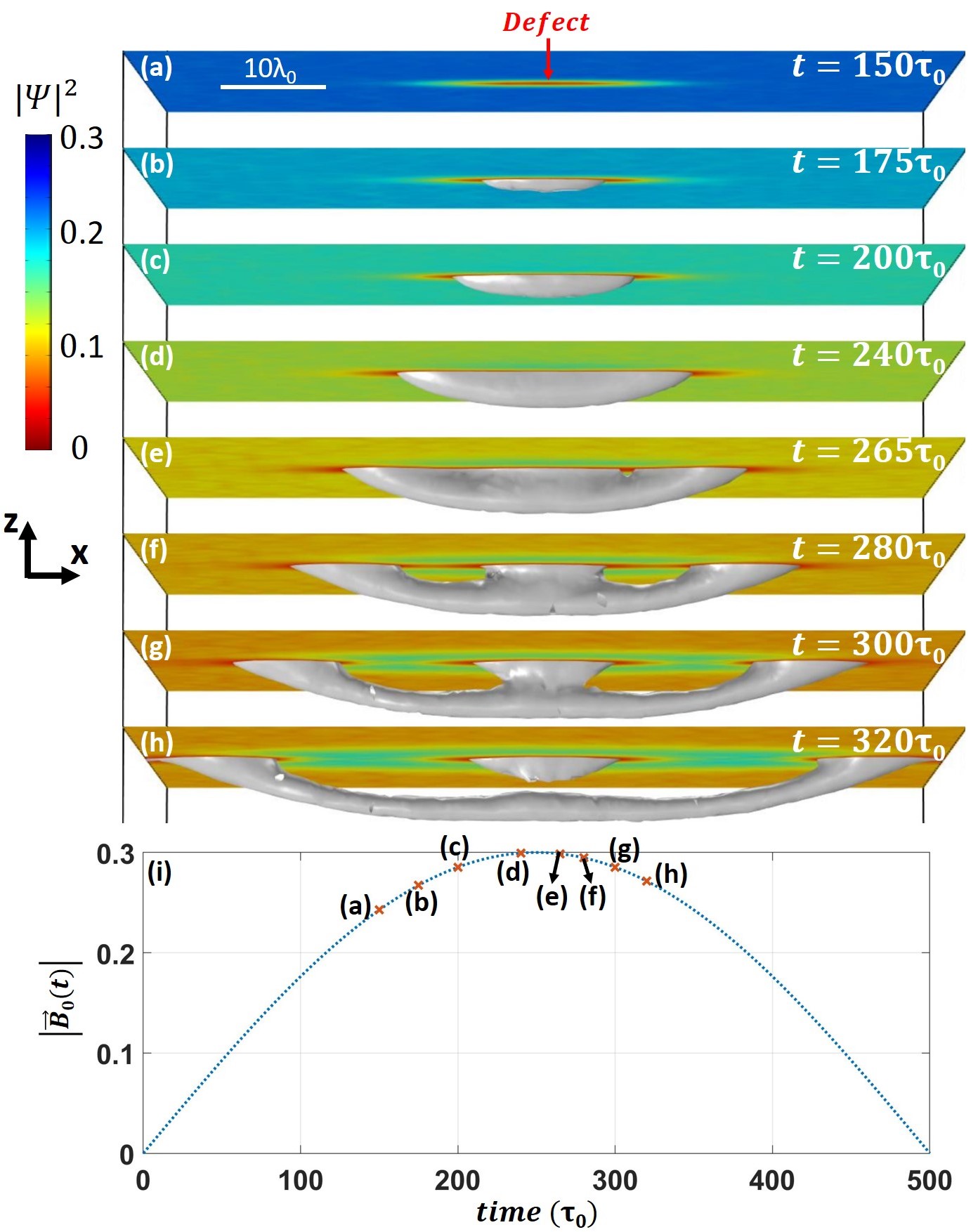}
\caption{(a)-(h) Plots of vortex semiloops illustrated with a silver surface (corresponding to $\left \vert \Psi \right \vert^2 =0.005$) at different times for parallel rf magnetic field in the $\hat{x}$ direction above the superconductor. A localized defect is placed at the origin ($\vec{r}_d=0\hat{x}+0\hat{y}+0\hat{z}$) with $\sigma_x=6$ and $\sigma_y=\sigma_z=1$ and $T_{cd}=0.1$. The color shows the order parameter magnitude $\left \vert \Psi \right \vert^2$ on the superconducting surface. (i) $\left \vert \vec{B_{0}} \right \vert$ at the surface vs time during the first half of the rf cycle. Red crosses correspond to field values for snapshots (a)-(h). The full list of simulation parameters is given in Table I. Note that this is a transient solution rather then a steady state solution.}
\label{Par_field}
\end{figure}

In previous sections, we examined the dynamics of vortex semiloops created by a point magnetic dipole, as it is relevant to the magnetic microscopy experiment \cite{Oripov2019}. In this section we will briefly address the more general case which is appropriate for SRF applications, a uniform parallel rf magnetic field $\left ( \vec{B}(t)=B_0 sin(\omega t) \hat{x} \right )$ above the superconductor in the presence of a single defect on the surface. In order to have truly uniform field, the boundary between superconductor and vacuum should be simulated as an infinite plane. To accurately simulate the screening currents on the surface of the cavity, the two-domain simulation method described in section IV.1 is used. The superconducting domain and vacuum domain are simulated inside 2 rectangular blocks instead of the cylindrical domain used in previous sections. The block dimensions are $L=80 \lambda_0$ (along the field direction) and width $W=60 \lambda_0$. The height of superconducting domain is $h_{sc} = 20 \lambda_0$, and the height of the vacuum domain is $h_{vac}=10 \lambda_0$. The vacuum domain is placed on the top of the superconducting domain. To mimic the infinite domain periodic boundary conditions are applied in the $\pm \hat{x}$ and $\pm \hat{y}$ directions both for $\Psi$ and $\vec{A}$. \par

Fig. \ref{Par_field} shows the solution for the order parameter in the case of externally applied rf magnetic field parallel to the surface of the superconductor along the $\hat{x}$ axis direction. A localized defect (modeled with Eq. \ref{DefectEq}) is placed at the origin ($\vec{r}_d=0\hat{x}+0\hat{y}+0\hat{z}$) with $\sigma_x=6$ and $\sigma_y=\sigma_z=1$ and $T_{cd}=0.1$. A \textit{transient} solution starting from the zero field Meissner state is studied in this case. A vortex semiloop penetrates into the superconducting domain at the site of the defect as the rf field amplitude increases \cite{Gurevich2015}. We consider vortex semiloops as a unique type of vortex, distinctly different from parallel line vortices \cite{Antoine2019}. When the amplitude of magnetic field is increased beyond that used in Fig.(\ref{Par_field}), we observe that arrays of parallel line vortices nucleate into the superconductor. While no defect was required to create vortex semiloops with the magnetic dipole source, a surface defect is required to create such a vortex when parallel field is applied. 

The solution shown in Fig. \ref{Par_field} is an initial transient solution, i.e. the simulation is not run for several cycles to reach the steady state condition. When the vortex semiloop reaches the boundary of the simulation in the field direction the results become nonphysical due to artificial pinning of the vortex semiloop by the boundaries. This finite size effect is currently limiting our ability to perform full rf parallel field simulation. Nevertheless, the transient solution shown in Fig. \ref{Par_field} may give some insight into the development of vortex semiloops in SRF cavities \cite{Gurevich2008}, and will be pursued in future work. 

\section{Discussion}
These simulations have proven very useful in understanding the measured third-harmonic response of Nb materials, subjected to intense localized rf magnetic fields \cite{Oripov2019}. In all the cases described in the previous section, the order parameter $\left \vert \Psi \right \vert^2$ and the vector potential $\vec{A}$ are first retrieved from the simulation. Using eq.(\ref{Currentdef}) the screening super-current is calculated for each point in space and time. The response magnetic field generated by said currents at the location of the dipole is calculated using the Biot-Savart law. The third-harmonic response recovered at the location of the dipole is obtained through Fourier transformation of the calculated response magnetic field. Later the TDGL-derived third-harmonic voltage $V_{3\omega}$ was compared with the third-harmonic response measured from experiment. The comparison is discussed in detail in Ref. \cite{Oripov2019}. \par

While most of the work was done for an oscillating parallel magnetic dipole, we also showed that vortex semiloops are created when a localized defect is introduced in the internal surface of an SRF cavity. It is plausible that vortex semiloops are one of the key sources of dissipation inside an SRF cavity at high operating power. The losses associated with such a vortex can be studied by combining the TDGL numerical technique with the experimental work published in \cite{Oripov2019}. \par

While recent advances in SRF cavity fabrication, especially the novel technique of Nitrogen doping and Nitrogen infusion \cite{Grassellino2013a, Martinello2016, Sauls2019}, have significantly improved the properties of SRF cavities, the microscopic mechanism responsible for this improvement is yet unknown. Nitrogen infused cavity surfaces can perhaps be thought of as a layered superconductor, with a dirty superconductor on top acting like a "slow" superconductor and suppressing vortex nucleation \cite{Transtrum2017, Romanenko2014a, RomanenkoIpac2018}. The characteristic time scale governing the dynamic behaviour of the superconductor was calculated by Gor'kov and Eliashberg to be $\tau_{GL}=\cfrac{\pi \hbar}{8 k_B (T_c-T)}$ \cite{Gorkov1968}. Superconductors in the dirty limit, with a finite inelastic electron-phonon scattering time $\tau_E$ subject to $\sqrt{D\tau_E}\ll\xi$ can be better studied using gTDGL, where the effect of a finite inelastic electron scattering time is considered \cite{Kramer1978}. However, Tinkham has argued that the characteristic time for the relaxation of the order parameter in a gapped superconductor in the clean limit should be much longer than the characteristic GL time $\tau_{GL}$, instead on the order of $\tau_E$ \cite{Scttmid1968, Tinkham2004Book, Kopnin2001book} ($\tau_E\approx 1.5 \times 10^{-10}$s for Nb \cite{Kaplan1976}). It has been argued that a TDGL-like equation that incorporates these long relaxation times, a so called slow-GL model, can be used in such circumstances \citep{Tinkham2004Book}. Perhaps the effects of Nitrogen doping and Nitrogen infusion on Nb cavities can be better understood by considering the effects of this different time scale on vortex semiloop formation. This can be accomplished with a sequence of TDGL, gTDGL and slow-GL model simulations. \par

There is also a proposal to create superconductor-insulator multilayer thin-film coatings with enhanced rf critical fields \cite{Gurevich2008Multilayer}. TDGL simulations can be used to guide the design process for these multilayers. Although TDGL is not a microscopic theory, and it is sometimes difficult to link the parameters of the model to observable experimental quantities, the general behaviour of the superconductor response to microwave magnetic fields, and the development of vortex semiloops, is still very insightful. \par

\begin{widetext}

\begin{table}[h]
\begin{ruledtabular}
\begin{tabular}{| Sc | Sc | Sc | Sc | Sc | Sc | Sc | Sc | Sc | Sc |}
\hline
 Parameter Name & \rotatebox[origin=c]{90}{Symbol} & \rotatebox[origin=c]{90}{Scale} & \rotatebox[origin=c]{90}{Fig. \ref{DipoleCompare}} & \rotatebox[origin=c]{90}{Fig. \ref{Vortex3D}} & \rotatebox[origin=c]{90}{Fig. \ref{DipolevsTime}} & \rotatebox[origin=c]{90}{Fig. \ref{DipoleExample_J_E}} & \rotatebox[origin=c]{90}{Fig. \ref{DipolevsField}} & \rotatebox[origin=c]{90}{Fig. \ref{Defect_Evol}} & \rotatebox[origin=c]{90}{Fig. \ref{Par_field}} \\ [0.5ex] 
 \hline\hline
 Temperature & $T$ & $T_c$ &0&0&0.6&0.9&0.6&0.85&0.7\\
 \hline
 \multirow{2}{*}{Applied RF field amplitude} & \multirow{2}{*}{$B_{0}$} & $B_{c2}$ &0.01&0.75&0.55&0.3&0.46-0.84&0.3&0.3\\
 
 \cline{3-10}
 & & $\mu_0H_{sh}^\dagger$ &0.009&0.69&0.49&0.268&0.41-0.75&0.268&0.268\\
 \hline
 Period of Applied RF field & $\cfrac{2 \pi}{\omega}$ & $\tau_0$ &Static&200&200&200&200&200&1000\\
 \hline
 Dipole height & $h_{dp}$ & $\lambda_0$ &1&8&8&12&8&12&-\\
 \hline
 Radius of the simulation domain & $R$ & $\lambda_0$ & 8 &12&35&60&20&40&80x60 \\
 \hline
 Height of superconducting domain & $\mu_0h_{sc}$ & $\lambda_0$ &10&6&20&50&8&25&20\\
 \hline
 Height of vacuum domain & $h_{vac}$ & $\lambda_0$ &5&3&20&25&4&15&10\\
 \hline
 Ginzburg-Landau parameter & $\kappa$&-&1&1&1&1&1&1&1 \\
 \hline
 Ratio of characteristic time scales & $\eta$ &-&1&1.675&1&0.2&1&0.5&1\\
 [1ex] 
 \hline
\end{tabular}
\end{ruledtabular}
\caption{Values of parameters used for TDGL simulations of the oscillating magnetic dipole above the superconductor. $^\dagger$ The static superheating field is calculated using the asymptotic formulas of reference \cite{Transtrum2011}. Here we use $\mu_oH_{sh}=0.9B_{c2}$. \label{ComsolTable}}
\end{table}

\end{widetext}

\section{Conclusion}
In this work we present a novel way to perform TDGL simulations in 3D for spatially nonuniform magnetic fields applied to a superconducting surface. Proof of principle results are presented to show the validity of the proposed two-domain simulation method. The vortex semiloops created by a point magnetic dipole above the surface, and the rf dynamics of such vortices are studied. The effect of temperature, rf field amplitude and the surface defects on the vortex semiloops are studied and presented. The resulting third-harmonic nonlinear response can be calculated and compared with the experimental data (comparison published in \cite{Oripov2019}). Finally, we demonstrate the creation of such rf vortex semiloops in the case of a uniform rf magnetic field parallel to a superconducting surface with a single defect.

\section*{Acknowledgments}
\begin{acknowledgments}
This research was conducted with support from U.S. Department of Energy/ High Energy Physics through Grant No.DESC0017931.
\end{acknowledgments}

\bibliography{library,citeTDGL}

\begin{thebibliography}{95}
\bibitem [{Vor({\natexlab{a}})}]{VortexSemiloopTimeAnimation}%
  \BibitemOpen
  \href@noop {} {} ({\natexlab{a}}),\ \bibinfo {note} {the time loop of the
  solution to the TDGL equations illustrated in Fig. 7 is available at 'Editors
  please paste link to VortexSemiloopTimeAnimation.gif'}\BibitemShut {NoStop}%
\bibitem [{Vor({\natexlab{b}})}]{VortexSemiloopFieldAnimation}%
  \BibitemOpen
  \href@noop {} {} ({\natexlab{b}}),\ \bibinfo {note} {solution to the TDGL
  equations as a function of peak applied magnetic field as illustrated in Fig.
  9 is available at 'Editors please paste link to
  VortexSemiloopFieldAnimation.gif'}\BibitemShut {NoStop}%
\end{thebibliography}

\end{document}